\documentclass[11pt]{article}
\usepackage[utf8]{inputenc}
\usepackage{amsmath}
\usepackage{graphicx}
\usepackage{enumerate}
\usepackage{natbib}
\usepackage[english]{babel}
\usepackage{amsfonts}
\usepackage{amsthm}
\graphicspath{ {Figures/} }
\usepackage{setspace}
\usepackage[table, dvipsnames]{xcolor}
\usepackage{comment}

\newcommand{\blind}{1}

\begin{document}

\def\spacingset#1{\renewcommand{\baselinestretch}%
{#1}\small\normalsize} \spacingset{1}


\if1\blind
{\title{\bf Optimal Designs for Two-Stage Inference}
  \author{Jonathan Stallrich\\ 
  North Carolina State University, Department of Statistics\\ 
  Michael McKibben\\
    North Carolina State University, Department of Statistics
    }
  \maketitle
} \fi

\if0\blind
{
  \bigskip
  \bigskip
  \bigskip
  \begin{center}
    {\LARGE\bf Optimal Designs for Two-Stage Inference}
\end{center}
  \medskip
} \fi

\bigskip
\begin{abstract}
\noindent 
The analysis of screening experiments is often done in two stages, starting with factor selection via an analysis under a main effects model. The success of this first stage is influenced by three components: (1) main effect estimators' variances and (2) bias, and (3) the estimate of the noise variance. Component (3) has only recently been given attention with design techniques that ensure an unbiased estimate of the noise variance. In this paper, we propose a design criterion based on expected confidence intervals of the first stage analysis that balances all three components. To address model misspecification, we propose a computationally-efficient all-subsets analysis and a corresponding constrained design criterion based on lack-of-fit. Scenarios found in existing design literature are revisited with our criteria and new designs are provided that improve upon existing methods.
\end{abstract}

\noindent
{\it Keywords: Pre-selection variance estimator, expected confidence interval criterion, all-subsets analysis, definitive screening design, pure error} 
\vfill

\newpage
\spacingset{1.45} 





\section{Introduction}\label{s:Intro}

Industries as varied as pharmaceuticals, manufacturing, and agriculture rely on sequential experimentation to optimize processes by varying the settings of one or more factors. The Pareto principle \citep{Box1986} states that roughly 20\% of the manipulated factors will account for 80\% of the variation of a response, so a pivotal first step is a screening experiment to identify these few important factors. 
Runs for screening experiments are often expensive, limiting the amount of information that can be gathered and thus complicating the analysis. To maximize the potential to identify the important factors, the experimenter must carefully pair the experimental design with a screening analysis strategy.

For each run of a designed experiment, the experimenter chooses values of the $k$ factors, denoted $\boldsymbol{d}^T=(d_1,\dots,d_k)$. For simplicity, numeric factors are scaled so $d_j \in [-1,1]$ and categorical factors are fixed at two values, where $d_j = \pm 1$. A design with $n$ runs is represented by the $n \times k$ matrix $\boldsymbol{D}$ with rows $\boldsymbol{d}_i^T=(d_{i1},\dots,d_{ik})$ and produces a corresponding $n \times 1$ response vector $\boldsymbol{y}$. The responses are taken to be generated by a linear model $\boldsymbol{y} = \boldsymbol{X}\boldsymbol{\beta} + \boldsymbol{e}$,
where $\boldsymbol{e}\sim N(\boldsymbol{0},\sigma^2\boldsymbol{I})$ and the $i$-th row of $\boldsymbol{X}$ is denoted $\boldsymbol{x}_i^T=(1,x_{i1},\dots,x_{ip})$ where $x_{il}$ is a function of one or more elements of $\boldsymbol{d}_i$ (the first element of $1$ is for the intercept effect, $\beta_0$).  The non-intercept elements of $\boldsymbol{\beta}$ measure a potential change in $\text{E}(y)$ due to changing the values of one or more of the factors. In practice, the necessary $x_{il}$'s are unknown so a common approach is to narrow the range of factor values so the true model can be approximated by main effects ($x_{il}=d_{ij}$), bilinear interaction effects ($x_{il}=d_{ij}d_{ij'}$), and, for numeric factors, quadratic effects ($x_{il}=d_{ij}^2$). 

We presume $\boldsymbol{\beta}$ can be partitioned into an active set, $\mathcal{A} = \{l > 0 \, : \, |\beta_l| > a\}$, and inert set, $\mathcal{I} = \{ l >0 \, : \, |\beta_l| \leq a\}$, where $a > 0$ is a multiple of $\sigma$. The corresponding set of active factors, $\mathcal{F}$, are those factors having at least one associated effect in $\mathcal{A}$. The primary screening inference goal is to accurately estimate $\mathcal{F}$ with a secondary goal of accurately estimating $\mathcal{A}$ and the values of the $\beta_l \in \mathcal{A}$. The factors deemed the most important may be further studied in later experiments to better achieve the secondary goal.



If the least-squares estimator for $\boldsymbol{\beta}$, denoted $\hat{\boldsymbol{\beta}}=(\boldsymbol{X}^T\boldsymbol{X})^{-1}\boldsymbol{X}^T\boldsymbol{y}$, is unique then we may estimate $\mathcal{A}$ by performing two-sided $t$-tests (with multiple testing adjustments if desired). The tests also require the following estimator for $\sigma^2$ to be well-defined
\begin{align}
\hat{\sigma}_X^2 = \frac{\boldsymbol{y}^T(\boldsymbol{I}-\boldsymbol{P}_X)\boldsymbol{y}}{n-\text{r}(\boldsymbol{X})}\ ,\ \label{eq:sigma2X}
\end{align}
where $\boldsymbol{P}_X=\boldsymbol{X}(\boldsymbol{X}^T\boldsymbol{X})^{-1}\boldsymbol{X}^T$ and $\text{r}(\boldsymbol{X})$ denotes the rank of $\boldsymbol{X}$. Henceforth, $\boldsymbol{P}_M=\boldsymbol{M}(\boldsymbol{M}^T\boldsymbol{M})^{-}\boldsymbol{M}^T$ denotes the projection matrix onto the column space of a matrix $\boldsymbol{M}$, which is well-defined even if $\boldsymbol{M}$ is not full rank. For a fixed $n$, the power of such a test is driven by the diagonal elements of $\text{Var}(\hat{\boldsymbol{\beta}})=\sigma^2 (\boldsymbol{X}^T \boldsymbol{X})^{-1}$. We will refer to $(\boldsymbol{X}^T \boldsymbol{X})^{-1}$ as the \emph{design variance matrix} and its diagonal elements as the \emph{design variances}. Denote the design variance of the $j$-th non-intercept estimator by $v_j$ and call $\sqrt{v_j}$ the design standard error. We can then rank all such designs having a correctly specified model using one of the ``alphabet" design criteria \citep{Kiefer}, such as the $D$- and $A$-criteria, that is a scalar function of $(\boldsymbol{X}^T\boldsymbol{X})^{-1}$.  An optimal design under such a criterion minimizes the design variances in some overall sense, maximizing the power while controlling for type 1 error, and hence improving our estimation of $\mathcal{F}$ and $\mathcal{A}$.


\subsection{ Two-Stage Analysis}

In practice, it is often infeasible to budget for $n > p+1$ runs. The principles of factor/effect sparsity, effect hierarchy, and effect heredity state that only a few $\beta_l$'s under such a model are in $\mathcal{A}$, and that main effects are more likely to be active than interactions and quadratic effects, referred to as second-order terms. This supports a two-stage analysis approach allowing for $n < p+1$ runs for screening: 
\begin{enumerate}
    \item Estimate a model with only main effects to screen all $k$ factors;
    \item Enforce effect heredity and investigate second-order effects of the important factors.
\end{enumerate}
This analysis strategy implies the partitioning $\boldsymbol{X} \boldsymbol{\beta}=     \boldsymbol{X}_1 \boldsymbol{\beta}_1 +  \boldsymbol{X}_2 \boldsymbol{\beta}_2$
where $\boldsymbol{X}_1=(\boldsymbol{1} \, | \, \boldsymbol{D})$ and $\boldsymbol{X}_2$ correspond to the second-order effects. Stage 1 estimates $\boldsymbol{\beta}_1$ with $\hat{\boldsymbol{\beta}}_1=(\boldsymbol{X}_1^T\boldsymbol{X}_1)^{-1}\boldsymbol{X}_1^T\boldsymbol{y}$ and $t$-tests are performed to estimate $\mathcal{F}$. An optimal design focused on this stage alone may be found by applying an alphabet criterion to $(\boldsymbol{X}_1^T\boldsymbol{X}_1)^{-1}$. Orthogonal main effect plans having $\boldsymbol{X}_1^T\boldsymbol{X}_1=n\boldsymbol{I}$ will be optimal for all such criteria, and include regular and nonregular fractional factorial designs (see Section~2). 

There are two complications to the stage 1 analysis. First, $\hat{\boldsymbol{\beta}}_1$ may be biased when $\boldsymbol{\beta}_2$ is nonzero. This bias may be partially measured by the elements of a design's alias matrix, $\boldsymbol{A}=(\boldsymbol{X}_1^T\boldsymbol{X}_1)^{-1}\boldsymbol{X}_1^T\boldsymbol{X}_2$. When $\boldsymbol{A}=\boldsymbol{0}$, $\hat{\boldsymbol{\beta}}_1$ will be unbiased regardless of $\boldsymbol{\beta}_2$. There are many criteria that jointly minimize the design variances of $\hat{\boldsymbol{\beta}}_1$ and the absolute magnitude of the $\boldsymbol{A}$. \cite{Jones_2011} proposed efficient designs with minimal aliasing (EDMA) minimizing $\text{tr}(\boldsymbol{A}^T\boldsymbol{A})$ subject to a $D$-efficiency constraint. Definitive Screening Designs (DSDs) from \cite{jones2011class} target two-stage screening of the full quadratic model with $d_j=\{0,\pm 1\}$, being both $D$-efficient and having $\boldsymbol{A}=\boldsymbol{0}$.



The second complication that has received less attention is that, $\hat{\sigma}^2_{X_1}$, the variance estimator based on $\boldsymbol{X}_1$, will be severely inflated unless $\boldsymbol{\beta}_2=\boldsymbol{0}$, leading to poor estimation of $\mathcal{F}$.
However, if $\text{r}(\boldsymbol{X})<n$ then $\hat{\sigma}_{X}^2$ is well-defined and unbiased for $\sigma^2$ and so may be used to perform testing of the main effects. We will refer to such an estimator as a \emph{pre-selection estimator} for $\sigma^2$ because it does not depend on the model selection procedure.  Suppose $\boldsymbol{X}$ has $n_u$ unique rows and the remaining $r=n-n_u$ rows are replicates. Denote the matrix of unique rows by $\boldsymbol{X}_u$. It is well-known that $\hat{\sigma}_{X}^2$ has the expression
\[
\hat{\sigma}_{X}^2=\frac{SS_{PE}+SS_{LOF}}{r+\ell}
\]
where $SS_{PE}$ refers to the pure-error sum of squares having $r$ degrees of freedom, and $SS_{LOF}$ refers to the lack-of-fit sum of squares having $\ell=n_u - \text{r}(\boldsymbol{X}_u)$ degrees of freedom. This estimator is a weighted average of the pure-error estimator, $SS_{PE}/r$, and what we call the lack-of-fit estimator, $SS_{LOF}/\ell$. The pure-error estimator is always unbiased, but the lack-of-fit estimator is unbiased only when the full model is correctly specified, i.e., $E(\boldsymbol{y})=\boldsymbol{X}\boldsymbol{\beta}$.

\cite{gilmour2012optimum} modified the $D$- and $A$-criterion to incorporate degrees of freedom of a pure-error estimator alone. They highlighted the need for such design criteria to incorporate $\sigma^2$ estimation due to concerns about model misspecification, but they ignore the potential bias of $\hat{\boldsymbol{\beta}}_1$ that would arise from model misspecification.  \cite{leonard2017bayesian} addressed the model misspecification concern with a criterion that can balance minimization of the design variances and biases, while requiring a number of replicated runs.  \cite{jones2017effective} proposed augmented DSDs (ADSDs) that add runs to a DSD to generate a lack-of-fit estimator while maintaining zero-aliasing. \cite{jones2020partial} considered including partial replication to ADSDs to produce a $\hat{\sigma}^2_X$ comprised of both pure-error and lack-of-fit degrees of freedom. A goal of this paper is to provide a unified design criterion that allows comparison of designs with any pre-selection estimator, regardless of how it comes about.



\subsection{ The Reactor Experiment}

Consider the $2^5$ full factorial reactor experiment from \cite{box1978statistics}. The resulting analysis gives the model:
\[
\hat{y} = 65.5 + 9.75d_{2}+5.375d_{4}-3.125d_5+6.625d_{2}d_{4} - 5.5 d_{4}d_{5}\ ,\
\]
making $\hat{\mathcal{F}}=\{2,4,5\}$. The estimated $\sigma^2$ after fitting the above model is $\hat{\sigma}^2=3.331^2$. \cite{Jones_2011} considered whether similar conclusions could be made from only 12 runs, comparing 3 potential designs: (a) a non-regular fractional factorial design (NRFFD), (b) a Bayesian $D$-optimal design \citep{dumouchel1994simple}, and (c) an EDMA design. Each design has 12 unique runs from the original $2^5$ design and are provided in the Supplementary Materials. We will compare these designs to a New Design shown in Table~\ref{tab:chap2_sect1_bestdesign} that replicates two runs and thus has 2 pure error degrees of freedom. To recreate the analysis for the New Design from an unreplicated $2^5$ design, we imposed hidden replication resulting from removing factors 1 and 3. For example, run 2 of our design is a replicate of run 1, and could take on one of the three remaining values of $y$ from the full factorial design where $(d_{i2},d_{i4},d_{i5})=(-1,-1,1)$. The same was done for run 4 of our design, giving nine possible analyses.  The potential values are reflected in Table~\ref{tab:chap2_sect1_bestdesign}.

\begin{table}[ht]
    \centering 
    \caption{New  12-run fractional factorial design. 
    Runs 2 and 4 each have 3 potential responses after projecting the $2^5$ full factorial experiment to factors $2, 4,$ and $5$. 
    }
    \begin{tabular}{c|cccccl|}
    \cline{2-7} 
         & 1 & 2 & 3 & 4 & 5 & $y$ \\ \cline{2-7}
         &      $+$ &   $-$ &   $-$ &    $-$ &    $+$  & $63$\\
         &      $+$ &   $-$ &   $-$ &    $-$ &    $+$  & $55, 56, 59$\\
         &      $-$ &   $+$ &   $+$ &    $+$ &    $-$  & $95$\\
         &      $-$ &   $+$ &   $+$ &    $+$ &    $-$  & $93, 94, 98$\\
         & $-$ &   $-$ &   $-$ &   $-$ &   $-$ & $61$\\
          & $+$ &    $+$ &    $+$ &    $+$ &    $+$ &  $82$\\
         & $-$ &   $-$ &   $-$ &   $+$ &    $+$ & $44$\\
         & $+$ &    $+$ &    $+$ &    $-$ &   $-$ & $61$\\
         & $+$ &   $-$ &    $+$ &    $+$ &   $-$ & $60$ \\
         & $-$ &    $+$ &   $-$ &   $-$ &    $+$ & $70$ \\
         
         & $-$ &   $-$ &    $+$ &   $-$ &    $+$ & $59$\\
         & $+$ &    $+$ &   $-$ &    $+$ &   $-$ & $93$\\
          \cline{2-7}
    \end{tabular}
\label{tab:chap2_sect1_bestdesign}
\end{table}

Table~\ref{tab:chap2_sect1_estimates} summarizes the designs' main effect estimates, design standard errors ($\sqrt{v_j}$), and aliasing with respect to a two-factor interaction model.  The reported estimates under the New Design are the average main effect estimates across the 9 potential analyses. Aliasing is summarized by $\sqrt{\boldsymbol{A}_j\boldsymbol{A}_j^T}$ where $\boldsymbol{A}_j$ is the $j$-th row of $\boldsymbol{A}$.  The NRFFD has the smallest design standard errors but largest aliasing. The Bayesian $D$-optimal design has a $D$-efficiency of $0.97$ and smaller but non-negligible aliasing. Both the EDMA and New Design have $\boldsymbol{A}=0$ with $D$-efficiencies of $0.92$ and $0.90$, respectively. The estimation biases for the NRFFD and Bayesian $D$-optimal designs are significant, increasing or decreasing the magnitude of many main effect estimates. 
The estimates under the EDMA and New Design are immune to these biases.


\begin{table}[ht]
    \centering
    \caption{Estimates of main effects ($\hat{\beta}_j$) under the main effect model, the design standard errors, and the aliasing values. The main effect estimates the New Design are the average estimates across its nine potential analyses.}
    \begin{tabular}{c c | ccccc}
    \multicolumn{2}{c}{}& \multicolumn{5}{c}{Factor}\\
    \multicolumn{2}{c}{}& 1 & 2 & 3 & 4 & \multicolumn{1}{c}{5}\\
    \hline 
        & Full Factorial & $-0.688$ & $\phantom{-}9.750$ & $-0.313$ & $\phantom{-}5.375$ & $-3.125$\\
        & NRFFD &  $-4.500$ & $\phantom{-}8.330$ & $-0.833$ & $\phantom{-}5.000$ & $-0.500$\\
    $\hat{\beta}_j$   & Bayes-$D$ & $-3.269$ & $\phantom{-}9.898$ & $-2.435$ & $\phantom{-}1.602$ & $-3.231$\\
        & EDMA & $\phantom{-}0.563$ & $\phantom{-}10.850$ & $-0.400$ & $\phantom{-}4.313$ & $-3.350$\\ 
        & New Design & $-0.694$ & $\phantom{-}10.597$ & $-0.403$ & $\phantom{-}3.847$ & $-2.847$\\
        \hline
        & NRFFD & $\phantom{-}0.289$ & $\phantom{-}0.289$ & $\phantom{-}0.289$ & $\phantom{-}0.289$ & $\phantom{-}0.289$ \\
    $\sqrt{v_j}$  &  Bayes-$D$ & $\phantom{-}0.293$ & $\phantom{-}0.293$ & $\phantom{-}0.293$ & $\phantom{-}0.293$ & $\phantom{-}0.293$\\
        & EDMA & $\phantom{-}0.306$ & $\phantom{-}0.316$ & $\phantom{-}0.316$ & $\phantom{-}0.306$ & $\phantom{-}0.316$ \\ 
        & New Design & $\phantom{-}0.289$ & $\phantom{-}0.323$ & $\phantom{-}0.323$ & $\phantom{-}0.323$ & $\phantom{-}0.323$\\
        \hline
           &  NRFFD &  $\phantom{-}0.816$ & $\phantom{-}0.816$ & $\phantom{-}0.816$ & $\phantom{-}0.816$ & $\phantom{-}0.816$\\
    $\sqrt{\boldsymbol{A}_j\boldsymbol{A}_j^T}$  &  Bayes-$D$ & $\phantom{-}0.531$ & $\phantom{-}0.531$ & $\phantom{-}0.531$ & $\phantom{-}0.531$ & $\phantom{-}0.531$\\
        & EDMA & 0 & 0 & 0 & 0 & 0\\ 
        & New Design & 0 & 0 & 0 & 0 & 0\\\hline
    \end{tabular}
\label{tab:chap2_sect1_estimates}
\end{table}

\begin{figure}[ht]
    \centering
    \includegraphics[width=0.4\textwidth]{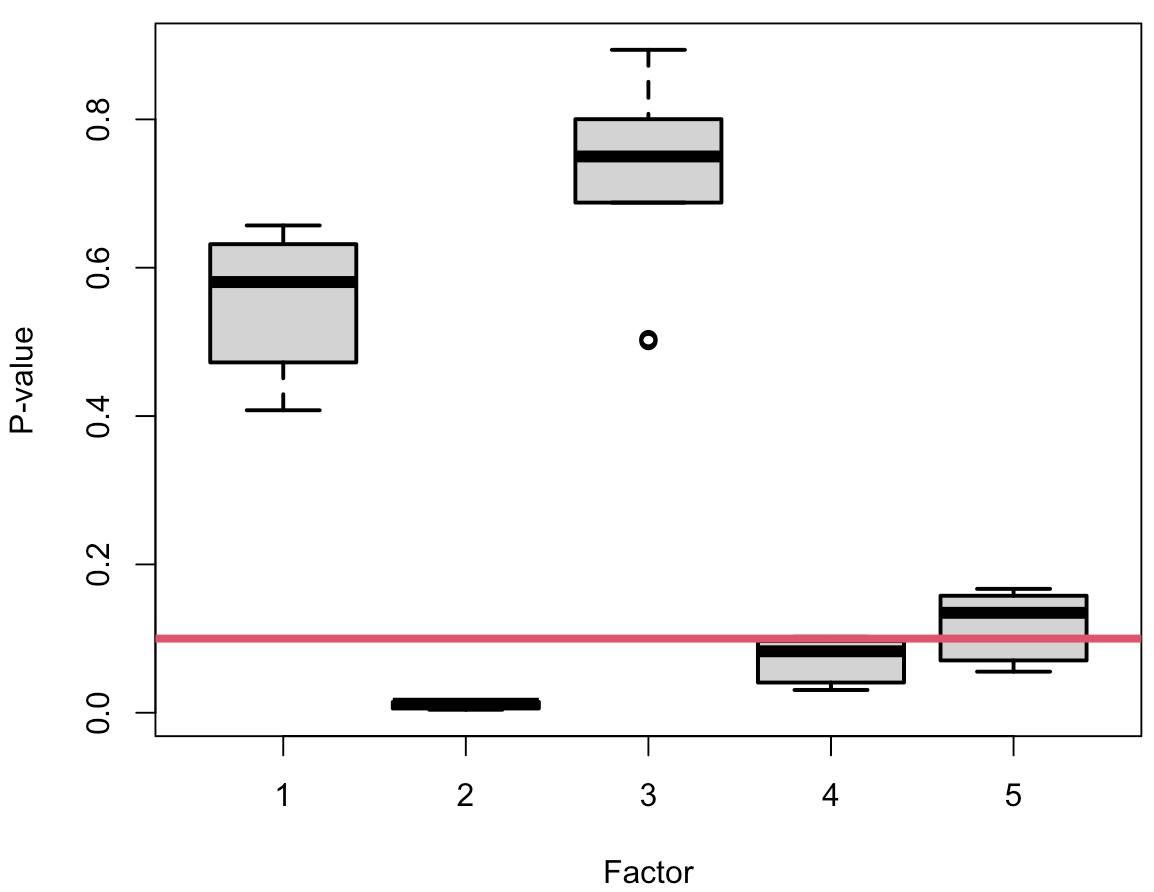}
    \caption{Boxplots of $p$-values for the main effects under the New Design's nine potential analyses.  The green squares refer to the $p$-values under the EDMA design. The red horizontal line visualizes a significance level of $\alpha=0.10$.}
\label{fig:chap2_sect1_pvals}
\end{figure}

\cite{Jones_2011} note that with a cutoff value of 3, screening under the EDMA design would match the screening of the full factorial design. However, this cutoff value is not based on a formal statistical test. We discovered the EDMA design has one degree of freedom from lack-of-fit under a two-factor interaction model, giving the unbiased estimate $\hat{\sigma}^2_{X} = 4.902^2$.  The New Design has two pure-error degrees of freedom, giving an average estimate (across the nine potential analyses) of $\hat{\sigma}^2_X=3.356^2$. While the EDMA's smaller $\sqrt{v_j}$ make it seem like a better design choice, the New Design's estimated standard errors, $\hat{\sigma}_X\sqrt{v_j}$, are smaller and these are what is used in the analysis. Moreover, having an extra degree of freedom for $\hat{\sigma}^2_X$ produces more powerful tests under the New Design. To see this, Figure~\ref{fig:chap2_sect1_pvals} shows the $p$-values for the main effect tests of the EDMA design and the New Design's nine potential analyses. For $\alpha=0.10$, only factor 2 would be deemed statistically significant under the EDMA design. For the New Design, every potential analysis correctly estimated factors 1 and 3 as inactive and factor 2 as active. Factor 4 was deemed active in all but one potential analysis while factor 5 was deemed active in 3 of the potential analyses. 
While neither design was able to perfectly recreate the full factorial results, the New Design's inferences were more accurate. This can be attributed to the New Design's ability to balance minimizing estimation variance and bias, as well as maximizing degrees of freedom for its pre-selection estimator of $\sigma^2$.




\subsection{ Contributions and Overview}

The first contribution of our paper is a criterion which jointly optimizes the aliasing, design variances, and degrees of freedom for a pre-selection estimator for $\sigma^2$, whether it be from pure-error or lack-of-fit. Using the equivalence between a two-sided $t$-test and checking whether a two-sided confidence interval includes 0, the process starts with optimizing a criterion that requires $\text{r}(\boldsymbol{X})< n$ and targets the end points of the two-sided confidence intervals for the main effect model using the pre-selection estimator $\hat{\sigma}^2_X$. 
The core idea is that reliable inference in the first stage is best achieved when the confidence intervals are narrow and centered at $\beta_j$. 
To construct designs under these criteria, we propose a modified coordinate exchange algorithm \citep{meyer1995coordinate} that considers simultaneous exchanges of replicated rows and emphasizes designs with lack-of-fit degrees of freedom. 

The proposed confidence interval criterion focuses on the first stage of the analysis and so may not perform well during the second stage analysis. The second contribution of our paper is an improvement upon the guided subsets analysis by \cite{jones2017effective} and \cite{weese2020strategies} with a computationally-efficient all-subsets analysis \citep{hofmann2020lmsubsets} in which model selection is performed by minimizing a modified Bayesian Information Criterion (BIC) that sets $\sigma^2=\hat{\sigma}^2_X$. To find a design that balances inference properties for both the first and second stage of the analysis, we adopt the constrained optimization approach by \cite{Jones_2011} and find a subset of designs generated under the confidence interval criterion according to a desired signal-to-noise detection threshold. We then rank these designs with a new reduced lack-of-fit criterion that measures the inflation of the modified BIC when important second-order terms are not included in the fitted model.

The paper is structured as follows.  Section~2 discusses relevant work to screening design criteria and unbiased estimators of $\sigma^2$.  Section~3 investigates the statistical properties of confidence intervals for the main effects and presents our proposed criterion for the first stage of the analysis. A coordinate exchange algorithm is described for optimizing the criterion. Section~4 
describes an all-subsets analysis as an alternative to the guided subsets analysis by \cite{jones2017effective}. Based on this analysis, a constrained criterion is presented to find a design with good model selection properties. In Section~5, we compare our computer-generated designs to different established screening designs, summarizing both their structural properties and comparing their performance in a simulation study. We conclude the paper with a discussion of our results and future work in Section~6.

\section{Background}\label{s:Background}

\subsection{Optimal Designs to Minimize Bias}

In the stage 1 analysis, $\text{E}(\hat{\boldsymbol{\beta}}_1)=\boldsymbol{\beta}_1 + \boldsymbol{A}\boldsymbol{\beta}_2$ and $\text{Var}(\hat{\boldsymbol{\beta}}_1)=\sigma^2 (\boldsymbol{X}_1^T\boldsymbol{X}_1)^{-1}$. Then $v_j$ denotes the $j$-th main effect design variance. We are interested in reducing bias by minimizing some measure applied to $\boldsymbol{A}$, while also minimizing the $v_j$. Such a strategy falls under the class of composite criteria \citep{pukelsheim2006optimal} that take a convex combination of the multiple criteria where weight $0 \leq w_c \leq 1$ is assigned to criterion $c$, such that $\sum_c w_c = 1$. It is unclear though how one should choose the weights so inevitably a range of values need to be considered followed by a further investigation to compare the trade-offs of each resulting optimal design. We now review the most popular relaxations to this approach.

For $2^k$ experiments, \emph{regular fractional factorial designs} (RFFDs) confound higher-order interaction effects with main effects and two-factor interactions. Motivated by effect hierarchy, several measures are able to evaluate the potential bias of main effects and two-factor interactions for an RFFD, with resolution and minimum aberration being the most popular \citep{cox2000theory}.  RFFDs require $n$ to be a power of $2$, while a \emph{non-regular fractional factorial design} (NRFFD) only requires that $n$ is a multiple of 4.  Both RFFDs and NRFFDs have $\boldsymbol{X}_1^T\boldsymbol{X}_1 = n \boldsymbol{I}$ but the aliasing under some NRFFDs can avoid perfectly confounding effects with one another \citep{mead2012statistical}. Criteria such as the generalized resolution and generalized minimum aberration \citep{GMA, Min_G_ab} criteria, and minimum $G_2$ aberration criterion \citep{G2} all extend resolution and minimum aberration to the more complicated aliasing structures of NRFFDs. Both RFFDs and NRFFDs emphasize minimizing design variances over bias.

A more balanced strategy to reduce design variances and bias is to express uncertainty about $\boldsymbol{\beta}_2$ by treating it as a random vector and assume it is drawn from some distribution independent of $\boldsymbol{\beta}_1$.  \cite{draper1992treating} assumed $\frac{1}{\sigma}\boldsymbol{\beta}_2 \sim N(\boldsymbol{0},\tau^2 \boldsymbol{I})$, making $\hat{\boldsymbol{\beta}}_1$ multivariate normal with mean $\boldsymbol{\beta}_1$ and variance $\sigma^2[(\boldsymbol{X}_1^T\boldsymbol{X}_1)^{-1} + \tau^2\boldsymbol{A}\boldsymbol{A}^T]$.  Applying an alphabet criterion to  $(\boldsymbol{X}_1^T\boldsymbol{X}_1)^{-1} + \tau^2\boldsymbol{A}\boldsymbol{A}^T$ should lead to an optimal design that balances variance and bias, depending on the scaling parameter $\tau^2$.  \cite{dumouchel1994simple} adopted a Bayesian interpretation of this approach, referring to $\boldsymbol{\beta}_1$ and $\boldsymbol{\beta}_2$ as primary and potential terms, respectively. The prior distribution for $\boldsymbol{\beta}_1$ is taken to be a flat, noninformative prior and again $\frac{1}{\sigma}\boldsymbol{\beta}_2 \sim N(\boldsymbol{0},\tau^2 \boldsymbol{I})$. Then the posterior covariance matrix for $\boldsymbol{\beta}$ is
$\sigma^2[\boldsymbol{X}^T \boldsymbol{X} + \tau^{-2}\boldsymbol{K}]^{-1}$
where $K$ is a diagonal matrix with $0$ elements corresponding to the primary terms $1$ for the remaining elements. They generate Bayesian $D$-optimal designs by applying the $D$-criterion to $\left(\boldsymbol{X}^T \boldsymbol{X} + \frac{1}{\tau^2}\boldsymbol{K}\right)^{-1}$. As $\tau \rightarrow \infty$, $\boldsymbol{X}^T \boldsymbol{X} + \frac{1}{\tau^2}\boldsymbol{K} \to \boldsymbol{X}^T \boldsymbol{X}$ so $\boldsymbol{X}$ must have full rank for the determinant to be nonzero. As $\tau \rightarrow 0$, the posterior variances for the potential terms will approach 0 regardless of the design and so the optimal design will focus on minimizing the posterior variances of the primary terms. 




EDMA's \citep{Jones_2011} were motivated by the constrained approach in \cite{MONTEPIEDRA199797} who minimized the mean-squared error for $\hat{\boldsymbol{\beta}}_1$, being $\textrm{Var}(\hat{\boldsymbol{\beta}}_1)+\boldsymbol{A}\boldsymbol{\beta}_2\boldsymbol{\beta}_2^T\boldsymbol{A}^T$, under a known $\boldsymbol{\beta}_2$. They define real-valued functions $\phi$ and $\Psi$ for $\boldsymbol{X}_1^T\boldsymbol{X}_1$ and $\boldsymbol{A}\boldsymbol{\beta}_2\boldsymbol{\beta}_2^T\boldsymbol{A}^T$, respectively. They then consider optimizing $\phi$  with a constraint on bias measure, and vice versa. 
Their constrained approach simplifies the difficult multi-objective optimization problem by declaring a cutoff value for one of the criteria. Such a cutoff implies that all designs whose criterion equal or improve the cutoff are equally valuable. We adopt this strategy in Section~4.2 using a signal-to-noise cutoff.

DSDs \citep{jones2011class}  balance minimizing main effect design variances and aliasing with respect to both two-factor interactions and quadratic effects. \cite{xiao2012} and \cite{jones2013definitive} showed that a DSD can be constructed by folding over a $k \times k$ conference matrix and appending a center run (i.e., $d_{ij}=0$). For an $n \times k$ design, $\tilde{\boldsymbol{D}}$, the foldover design is $\boldsymbol{D}^T=(\tilde{\boldsymbol{D}}^T \, | \, -\tilde{\boldsymbol{D}})$,
having $2n$ runs. This technique is popular when constructing a design with $\boldsymbol{A}=\boldsymbol{0}$ for the two-factor interaction model. 

\subsection{Optimal Designs for Model Selection}\label{sec:ModelSelection}
Designs that minimize aliasing may still struggle with identifying the important second-order terms, which may be done with some stepwise method \citep{Efroymson}.  
Stepwise methods efficiently generate many plausible models and choose the best model that minimizes a model selection criterion such as the Akaike information criterion (AIC), corrected AIC (AICc) \citep{AICc}, and the Bayesian Information Criterion, or BIC \citep{BIC}. For a general linear model with $E(\boldsymbol{y})=\boldsymbol{X}\boldsymbol{\beta}$, BIC is
\begin{align*}
	\text{BIC}&=-2\log(\hat{L}) = n \log\left(\frac{\boldsymbol{y}^T(\boldsymbol{I}-\boldsymbol{P}_X)\boldsymbol{y}}{n}\right) +c \ \text{log}(n),    
	\end{align*}
where $c$ is the number of estimated parameters in the selected model, which includes $\sigma^2$. 
For BIC to reliably identify the correct model, a number of properties must hold: 
\begin{enumerate}
    \item The design should be able to estimate as many unique models as possible;
    \item For an incorrectly specified model, $\boldsymbol{y}^T(\boldsymbol{I}-\boldsymbol{P}_X)\boldsymbol{y}$ should be inflated;
    \item For two model matrices $\boldsymbol{X}$ and $\tilde{\boldsymbol{X}}$ that correctly specify the model where the columns of $\boldsymbol{X}$ are included in $\tilde{\boldsymbol{X}}$, $\boldsymbol{y}^T(\boldsymbol{I}-\boldsymbol{P}_X)\boldsymbol{y} \approx \boldsymbol{y}^T(\boldsymbol{I}-\boldsymbol{P}_{\tilde{X}})\boldsymbol{y}$.
\end{enumerate}
Conditions (1) and (2) have motivated criteria for screening experiments that hone in on designs that can uniquely estimate and differentiate different models. Examples include the Estimation Capacity \citep{Sun}, Information Capacity \citep{MRFD}, and the $Q_B$ criterion \citep{qb}. One shortcoming of these criteria is that they do not guarantee that it is possible to differentiate between two competing models, leading to the criteria Subspace Angle (SA), Maximum Prediction Difference (MPD), and Expected Prediction Difference (EPD) proposed by \cite{JLN}. Condition (3) is difficult to satisfy because $\boldsymbol{y}^T(\boldsymbol{I}-\boldsymbol{P}_X)\boldsymbol{y} \geq \boldsymbol{y}^T(\boldsymbol{I}-\boldsymbol{P}_{\tilde{X}})\boldsymbol{y}$ and the difference increases as more columns are added to $\tilde{\boldsymbol{X}}$. This difference should be offset by an increase in the function on $c$, but there is no guarantee this will happen. Indeed, $\log(\boldsymbol{y}^T(\boldsymbol{I}-\boldsymbol{P}_{\tilde{X}})\boldsymbol{y}/n)$ can quickly approach $-\infty$, exacerbating the issue.  In Section~4.1, we show how a modified BIC incorporating a pre-selection estimator can remedy this issue and in Section 4.2 we propose a design criterion that measures the expected inflation of this modified BIC across many second-order models.

\subsection{Optimal Designs Requiring Variance Estimation}

The designs in Section~2.2 do not guarantee accurate estimation of $\sigma^2$ needed to perform inference in the first stage. \cite{Lu2011} addressed this with a Pareto front approach involving three screening design criteria: (1) $|\boldsymbol{X}_1^T\boldsymbol{X}_1|$, (2) $\text{tr}(\boldsymbol{A}\boldsymbol{A}^T)$, and (3) a function of bias of $\hat{\sigma}^2_{X_1}$. While their approach has many benefits, $\hat{\sigma}^2_{X_1}$ will almost always be an inflated estimator of $\sigma^2$.



\cite{gilmour2012optimum} proposed criteria requiring $r$ replicated runs to produce a pure-error estimator for $\sigma^2$. Let $F_{k,r,1-\alpha}$ denote the quantile corresponding to probability $1-\alpha$ of an $F$-distribution with $k$ and $r$ numerator and denominator degrees of freedom, respectively.  Let $\boldsymbol{P}_1=\frac{1}{n}\boldsymbol{1}\boldsymbol{1}^T$, making $[\boldsymbol{D}^T(\boldsymbol{I}-\boldsymbol{P}_1)\boldsymbol{D}]^{-1}$ the design variance matrix for the main effects adjusted for the intercept. For the main effect analysis, their modified $D$-criterion is
\begin{align}
 \frac{(F_{k,r,1-\alpha})^{k}}{|\boldsymbol{D}^T(\boldsymbol{I}-\boldsymbol{P}_1)\boldsymbol{D}|}\ ,\ \label{e:GTDcrit}
 \end{align}
and their modified $A$-criterion is
\begin{align}
  \ F_{1,r,1-\alpha}\text{tr}[(\boldsymbol{D}^T(\boldsymbol{I}-\boldsymbol{P}_1)\boldsymbol{D})^{-1}]\ .\
  \label{e:GTAcrit}
 \end{align}
Criterion \eqref{e:GTAcrit} is equivalent to minimizing the margins of error of a confidence interval with $\hat{\sigma}^2$ set to $\sigma^2$.  For fixed $r$, the $F$-quantiles force a constraint on the types of designs one may consider (i.e., those with $r$ replicates). Including the $F$-quantiles will be useful to compare designs with varying $r$, although the impact of increasing $r$ depends on the chosen $\alpha$. The criteria of \cite{gilmour2012optimum} ignore the potential impact of model misspecification on estimation bias. \cite{leonard2017bayesian} balance minimizing design variances and biases of the main effect model, while requiring unbiased variance estimation with a Bayesian version of \eqref{e:GTDcrit}. They then used the techniques described in \cite{Lu2011} to further compare their generated designs. 


An ADSD \citep{jones2017effective} generates a pre-selection estimator of $\sigma^2$ based on $f$ fake factors added to a DSD. For $k$ factors, they construct a DSD with $k+f$ giving a design with $2(k+f)+1$ runs in which the $k+f$ main effect columns are mutually orthogonal and orthogonal to all two-factor interactions and quadratic effects. Therefore, the sum of squares from the $f$ fake factors can be used to create a pre-selection estimator with $f$ degrees of freedom.
\cite{jones2020partial} further considered ADSDs with replicated runs, but found the lack-of-fit approach tended to have smaller design variances. 

\section{Expected confidence interval criterion}\label{s:Theory}

When $n-\text{r}(\boldsymbol{X})=g \geq 1$ there exists a pre-selection estimator $\hat{\sigma}_X^2$ with $g$ degrees of freedom. A two-sided $t$-test for the $j$-th main effect analysis involves calculating the $t$-statistic $T=\hat{\beta}_j/\sqrt{\hat{\sigma}_X^2 v_j}$
and checking whether $|T| > t_{\alpha/2,g}$. An equivalent procedure is to calculate the $100(1-\alpha)$\% confidence interval
\[
\hat{\beta}_j \pm t_{\alpha/2,g}\sqrt{\hat{\sigma}^2_X v_j}
\]
and check whether the interval contains 0.  The width of the confidence interval is controlled by the margin of error $t_{\alpha/2,g}\sqrt{\hat{\sigma}^2_X v_j}$, which is a random quantity even for a fixed design. The center of the confidence interval is $\hat{\beta}_j$ and so is also a random quantity. The margin of error compensates for the variance of $\hat{\beta}_j$ so that the probability a randomly generated interval captures $\beta_j$ is $1-\alpha$. The margin of error, however, does not address the potential bias of $\hat{\beta}_j$.  It is possible for the bias to pull a randomly generated interval for $\beta_j \in \mathcal{A}$ towards zero, decreasing power. On the other hand, the bias may unintentionally improve estimation of $\mathcal{A}$ by pushing intervals for $\beta_j \in \mathcal{A}$ further away from 0. For $\beta_l \in \mathcal{I}$ the bias will always push intervals away from zero, thereby increasing the type 1 error rate over the desired $\alpha$. From a design perspective, we can improve inference based on confidence intervals by making the intervals accurate (i.e., centered at $\beta_j$) and precise (i.e., narrow margins of error).


Define a design's expected confidence interval for the $j$-th main effect as
\[
(\beta_j + \boldsymbol{A}_j\boldsymbol{\beta}_2) \pm E\left(\sqrt{\hat{\sigma}^2_X}\right) t_{\alpha/2,g} \sqrt{v_j} \ .\
\]
We remove dependence of the unknown $\sigma^2$ by dividing the endpoints of the interval by $\sigma$, yielding an interval for $\beta_j/\sigma$, being the main effect's signal-to-noise ratio. In the Supplementary Materials, we derive $\frac{1}{\sigma}E(\sqrt{\hat{\sigma}_X^2})$ and plug this into the confidence interval expression.
For $c(\alpha,g)=\frac{1}{\sigma}E(\sqrt{\hat{\sigma}_X^2})t_{\alpha/2,g}$, it follows that the maximum deviation of the expected confidence interval from $\beta_j/\sigma$ is $\frac{1}{\sigma}|\boldsymbol{A}_j\boldsymbol{\beta}_2|+c(\alpha,g)\sqrt{v_j}$ and that minimizing this value will give an accurate and precise interval. Therefore, we desire a design that minimizes the average of these deviations:
\[
\textrm{min} \frac{1}{k} \sum_{j=1}^k \frac{1}{\sigma}|\boldsymbol{A}_j\boldsymbol{\beta}_2| + c(\alpha,g) \sqrt{v_j}\ .\
\]
This approach maps a complicated, multi-criteria optimization problem into a single, interpretable value. It informs us of a design's ability to screen main effects with a given signal-to-noise ratio. For example, if the criterion value is 1 and $\boldsymbol{A}=0$, then, on average, we should consistently detect active effects whose signal-to-noise ratio is 1 or higher with the desired $\alpha$ level. This interpretation will be useful for our constrained optimization approach described in Section~4.2.

To evaluate the criterion, one must specify a value for $\frac{1}{\sigma}\boldsymbol{\beta}_2$ or posit a distribution. We assume $\frac{1}{\sigma}\boldsymbol{\beta}_2 \sim N(0,\tau^2 \boldsymbol{I})$ and consider the expected value of the criterion with respect to this distribution.  In the Supplementary Materials, we show that $\frac{1}{\sigma}|\boldsymbol{A}_j\boldsymbol{\beta}_2|$ follows a half-normal distribution with
\[
E\left(\frac{1}{\sigma}|\boldsymbol{A}_j\boldsymbol{\beta}_2|\right)=\sqrt{\frac{2\tau^2}{\pi}\boldsymbol{A}_j\boldsymbol{A}_j^T}\ .\
\]
Our proposed \emph{expected confidence interval} (ECI) criterion, is then
\begin{align}
\textrm{min} \ \frac{1}{k} \sum_{j=1}^k \left[\sqrt{\frac{2\tau^2}{\pi}\boldsymbol{A}_j\boldsymbol{A}_j^T} + c(\alpha,g) \sqrt{v_j}\right]\ .\ \label{eqn:ECInobias}
\end{align}
As $\tau^2 \to 0$, the ECI criterion focuses on minimization of $c(\alpha,g)\sum_l \sqrt{v_j}$, which is similar to \cite{gilmour2012optimum} except it does not require only pure error degrees of freedom. As $\tau^2 \to \infty$, the criterion will focus more attention on minimizing bias.



 Identifying a design that minimizes \eqref{eqn:ECInobias} is challenging because it is not clear whether the $g$ degrees of freedom should come from pure-error, lack-of-fit, or some combination of the two. Moreover, while generating a design with pure-error degrees of freedom simply requires replicating rows of $\boldsymbol{D}$, we are unaware of a general technique for creating lack-of-fit degrees of freedom. 
We propose a design construction algorithm for which the user specifies $g$ through a minimum number of desired pure-error degrees of freedom, $r$, and lack-of-fit degrees of freedom, $\ell$.  The algorithm ensures the pure-error condition is met via replication, while modifying \eqref{eqn:ECInobias} to penalize designs that do not have at least $\ell$ lack-of-fit degrees of freedom. To understand this modification, assume $\boldsymbol{X}$ has no replicated runs (i.e., $\boldsymbol{X}=\boldsymbol{X}_u$); otherwise the following derivations apply to $\boldsymbol{X}_u$. If $\text{r}(\boldsymbol{I}-\boldsymbol{P}_X) = \tilde{\ell} \geq \ell$, then we have met the minimum desired lack-of-fit degrees of freedom and evaluate the design under \eqref{eqn:ECInobias}. Otherwise, we construct a lack-of-fit estimator $\hat{\sigma}^2_{L}=\boldsymbol{y}^T\boldsymbol{P}_L\boldsymbol{y}/\ell$
from a full-rank $n \times \ell$ matrix, $\boldsymbol{L}$, that estimates $\sigma^2$ with minimal bias. Requiring $\boldsymbol{X}_1^T\boldsymbol{L}=0$ and taking expectation with respect to the prior on $\frac{1}{\sigma}\boldsymbol{\beta}_2$, we show in the Supplementary Materials that


\[
\text{E}\left(\frac{\hat{\sigma}^2_{L}}{\sigma^2}\right)=1 + \frac{\tau^2}{\ell}\text{tr}(\boldsymbol{X}_2^T\boldsymbol{P}_{L}\boldsymbol{X}_2)\ .\ 
\]
 The bias of $\hat{\sigma}^2_{L}$ depends on $\tau^2$ and the orthogonality between $\boldsymbol{L}$ and $\boldsymbol{X}_2$, with the ideal case of $\boldsymbol{X}_2^T\boldsymbol{L}=0$ which is possible if and only if $\text{r}(\boldsymbol{I}-\boldsymbol{P}_X) \geq \ell$. If $\text{r}(\boldsymbol{I}-\boldsymbol{P}_X)=\tilde{\ell}<\ell$ then $\tilde{\ell}$ columns of $\boldsymbol{L}$ should come from the null space of $\boldsymbol{X}$, as they will be orthogonal to $\boldsymbol{X}_1$ and $\boldsymbol{X}_2$, leaving $\ell^*=\ell-\tilde{\ell}$ further columns to compute $\hat{\sigma}^2_{L}$. These columns, denoted $\boldsymbol{L}^*$, must be constructed from the column space of 
$\boldsymbol{X}_{2|1}=(\boldsymbol{I}-\boldsymbol{P}_{X_1})\boldsymbol{X}_2$, i.e., $\boldsymbol{L}^*=\boldsymbol{X}_{2|1}\boldsymbol{M}$ for some $\boldsymbol{M}$. Then  $\text{tr}(\boldsymbol{X}_2^T\boldsymbol{P}_{L}\boldsymbol{X}_2)=\text{tr}(\boldsymbol{M}^T\boldsymbol{C}_{2|1}^2\boldsymbol{M}\left[\boldsymbol{M}^T\boldsymbol{C}_{2|1}\boldsymbol{M}\right]^{-1})$, where $\boldsymbol{C}_{2|1}=\boldsymbol{X}_{2|1}^T\boldsymbol{X}_{2|1}$. To minimize the bias of $\hat{\sigma}^2_{L}$, we let the columns of $\boldsymbol{M}$ be the $\ell^*$ eigenvectors of $\boldsymbol{C}_{2|1}$ corresponding to the smallest positive eigenvalues, making $\text{tr}(\boldsymbol{X}_2^T\boldsymbol{P}_{L}\boldsymbol{X}_2) = \sum_{m=1}^{\ell^*} \lambda_m$.

When $\tilde{\ell}<\ell$, we need to correct $E(\sqrt{\hat{\sigma}_X^2})$ that appears in $c(\alpha,g)$ to account for this bias. It is difficult to derive $E(\sqrt{\hat{\sigma}_X^2})$ when $\tilde{\ell} < \ell$ so we use its upper bound, $\sqrt{E(\hat{\sigma}_X^2)}$, thereby replacing $c(\alpha,g)$ with $$c(\alpha,g,\tau)=t_{\alpha/2,g}\sqrt{1+\frac{\tau^2}{g}\sum_{m=1}^{\ell^*} \lambda_m}\ .\ $$ The denominator of $g=r+\ell$ rather than $\ell$ is due to our consideration of replicated runs (see Supplementary Materials). As $\tau^2 \to 0$ the updated ECI criterion again focuses on $t_{\alpha/2,g}\sum_j \sqrt{v_j}$ but as $\tau^2 \to \infty$ and $\ell>0$ the $c(\alpha,g,\tau)$ can grow large when $\tilde{\ell} < \ell$, causing the construction algorithm to highlight designs with $\tilde{\ell}\geq \ell$.






The design search algorithm is described in the Supplementary Materials for a fixed $\tau$, minimum number of replicated runs, $r$, and lack-of-fit degrees of freedom, $\ell$. To summarize, we represent the design matrix as $\boldsymbol{D}^T=(\boldsymbol{D}_u^T\, | \, \boldsymbol{D}_r^T)$ where $\boldsymbol{D}_u$ are the unrestricted, but not necessarily unique, design settings and $\boldsymbol{D}_r$ is comprised of the minimum $r$ replicated runs the repeat one or more rows in $\boldsymbol{D}_u$. Each row in $\boldsymbol{D}_r$ is then paired with a row in $\boldsymbol{D}_u$, and we allow multiple rows in $\boldsymbol{D}_r$ to be paired to a single row in $\boldsymbol{D}_u$. It is also possible for $\boldsymbol{D}_u$ to contain more replicated rows. For a given initial design matrix, the proposed algorithm iterates between a coordinate exchange algorithm of $\boldsymbol{D}_u$ and all-subsets optimization of $\boldsymbol{D}_r$ using the rows of the current $\boldsymbol{D}_u$. 


\section{An all-subsets analysis and constrained design criterion}

The ECI criterion targets main effect screening and so may sacrifice inference properties of the second-stage analysis.  That is, the design may be overpowered for main effect screening, especially if the active main effects have a large signal-to-noise ratio. Following \cite{Jones_2011}, we propose a  constrained criterion that ranks designs achieving an ECI criterion value related to a desired detectable signal-to-noise ratio of the main effects.  The constrained criterion is motivated by a modification to the guided subsets analysis from \cite{jones2017effective}, which first performs inference on the main effect model and pools the sum of squares from the inactive main effects with $\boldsymbol{y}^T(\boldsymbol{I}-\boldsymbol{P}_X)\boldsymbol{y}$, giving an updated estimator for $\sigma^2$ with $g^*\geq g$ degrees of freedom. For simplicity, we will refer to this quantity by simply $\hat{\sigma}^2$. The guided subsets approach allows for different types of effect heredity, which determines the second-order models considered:
\begin{enumerate}
    \item Strong heredity: only second-order effects involving factors in $\hat{\mathcal{F}}$;
    \item Weak heredity: only second-order effects involving at least one factor in $\hat{\mathcal{F}}$;
    \item Full: All second-order effects.
\end{enumerate}
For simplicity of notation, let $\boldsymbol{X}_2$ denote the columns of all second-order effects under consideration, defined by $\boldsymbol{\beta}_2$. For $\boldsymbol{X}_{2|1}=(\boldsymbol{I}-\boldsymbol{P}_{X_1})\boldsymbol{X}_2$, the guided subsets approach starts with an overall test on the second-order effects of interest with $F$-statistic
\[
F=\frac{\boldsymbol{y}^T\boldsymbol{P}_{X_{2|1}}\boldsymbol{y}/\text{r}(\boldsymbol{X}_{2|1})}{\hat{\sigma}^2}\ ,\
\]
which under $H_0: \boldsymbol{\beta}_2=\boldsymbol{0}$ follows an $F$-distribution with $\text{r}(\boldsymbol{X}_{2|1})$ and $g^*$ numerator and denominator degrees of freedom, respectively.  This expression for the $F$-statistic is a generalized version of the one in \cite{jones2017effective}, as their expression was written for DSDs without replication. More details are provided in the Supplementary Materials.


If $H_0$ is rejected, the guided subsets approach systematically assesses the  residual sum of squares across models including only a subset of the second-order terms to determine if any potential terms among those in $\boldsymbol{X}_2$ have been left out. This is equivalent to a reduced lack-of-fit test where the largest possible model is assumed to satisfy $E(\boldsymbol{y})=\boldsymbol{X}\boldsymbol{\beta}$. Let $\boldsymbol{Z}_{2|1}$ denote the $p_2$ columns from $\boldsymbol{X}_{2|1}$ for the proposed model. Requiring $\boldsymbol{Z}_{2|1}$ to have full column rank, the reduced lack-of-fit sum of squares is $\boldsymbol{y}^T(\boldsymbol{P}_{X_{2|1}}-\boldsymbol{P}_{Z_{2|1}})\boldsymbol{y}$ and the test statistic is
\[
F_{Z_{2|1}}=\frac{\boldsymbol{y}^T(\boldsymbol{P}_{X_{2|1}}-\boldsymbol{P}_{Z_{2|1}})\boldsymbol{y}/(\text{r}(\boldsymbol{X}_{2|1})-p_2)}{\hat{\sigma}^2}\ .\
\]
Using $\alpha=0.20$, if $F_{Z_{2|1}}\leq F_{0.2,\text{r}(\boldsymbol{X}_{2|1})-p_2,g^*}$ the model is determined to not exhibit reduced lack-of-fit, providing evidence that the fitted model includes all important second-order effects. The guided subsets approach starts with consideration of all models including one second-order effect. If all such models have $F_{Z_{2|1}}>F_{0.2,\text{r}(\boldsymbol{X}_{2|1})-p_2,g^*}$, all models including two second-order effects are considered. This process continues until a $\boldsymbol{Z}_{2|1}$ is found that exhibits no reduced lack-of-fit but only up to models with $\text{r}(\boldsymbol{X}_{2|1})/2$ terms. The justification for considering only models up to this size was based on an investigation by \cite{jones2017effective} on the model discrimination capabilities of ADSDs. We could not find a recommendation for the scenario in which all models of this size exhibit lack-of-fit, nor did we find any investigation about what would happen if this bound increased.

While intuitive, there are some potential issues with the guided subsets analysis. If multiple models of a given size do not exhibit reduced lack-of-fit, \cite{jones2017effective} recommend that the model with the smallest $\boldsymbol{y}^T(\boldsymbol{P}_{X_{2|1}}-\boldsymbol{P}_{Z_{2|1}})\boldsymbol{y}$ be selected. \cite{weese2021strategies} noted that this leads to a random selection among the models that do not exhibit reduced lack-of-fit and recommended instead that the terms from all such models be pooled together to generate a list of potentially important terms. This leads to increased ability to identify active second-order effects, but will also lead to a higher false discovery rate. The guided subsets approach tends to have a negligible false discovery rate but is prone to missing active second-order effects.

\subsection{All-subsets analysis}

To address the issues with the guided subsets analysis, we propose all-subsets regression to compare a larger collection of models and choose the best model according to a  modified BIC that conditions on $\sigma^2 = \hat{\sigma}^2$: 
\[
    \text{mBIC}=\frac{\boldsymbol{y}^T(\boldsymbol{I}-\boldsymbol{P}_{X_1}-\boldsymbol{P}_{Z_{2|1}})\boldsymbol{y}}{\hat{\sigma}^2} + \log(n)(1+k+p_2)\ ,\
\]
where $\boldsymbol{y}^T(\boldsymbol{I}-\boldsymbol{P}_{X_1}-\boldsymbol{P}_{Z_{2|1}})\boldsymbol{y}$ is the residual sum of squares for the main effect model (perhaps reduced to the set of important factors from stage 1) plus $p_2$ second-order terms in $\boldsymbol{Z}_{2|1}$.
See \cite{ESLtextbook} for details on the derivation. Note mBIC does not include the $\log$ function which partially remedies issue (3) in Section~\ref{sec:ModelSelection}. Using the \texttt{R} package \texttt{lmSubsets} \citep{hofmann2020lmsubsets}, we found the calculation of all residual sum of squares to be surprisingly fast even for $k=10$. For exceptionally large model spaces, one could set an upper bound on the model size considered. Just like with guided subsets, we can also focus attention on certain submodels according to a heredity rule, but there is no need to specify a cutoff value; the model with the smallest BIC produces the estimate for $\mathcal{A}$. 

In the Supplementary Materials, we perform a simulation study to compare the all-subsets mBIC analysis to two versions of the guided subsets analysis, one with a maximal model size of $\lfloor \text{r}(\boldsymbol{X}_2)/2 \rfloor$ (as recommended by the \cite{jones2017effective}) and $\text{r}(\boldsymbol{X}_2)-1$. All versions assume strong heredity when compiling the
potential second-order effects. We focus on ADSDs with $k=6$ and $k=8$ and $f=2$, which gives two lack-of-fit degrees of freedom.  As expected, guided subsets with a maximal model size of $\lfloor \text{r}(\boldsymbol{X}_2)/2 \rfloor$ had the lowest false positive rates while all-subsets had the highest false positive rates. Generally the higher false positive rate was accompanied by a larger true positive rate, which is a reasonable compromise given how few runs are performed. We would likely get similar results from the guided subsets with $\text{r}(\boldsymbol{X}_2)-1$ by increasing $\alpha$ but it is not clear how to pick the value. Again, a major benefit of the all-subsets is that it does not require specification of a cutoff. 

\subsection{A constrained, reduced lack-of-fit (rLOF) criterion}

 Conditions (1) and (2) from Section~\ref{sec:ModelSelection} imply the rank of $\boldsymbol{X}_{2|1}$ should be as large as possible and that $\boldsymbol{y}^T(\boldsymbol{I}-\boldsymbol{P}_{X_1}-\boldsymbol{P}_{Z_{2|1}})\boldsymbol{y}$ should be inflated if important second-order terms are not included. In the Supplementary Materials, we show that minimizing mBIC across all second-order models is equivalent to minimizing
\[
\frac{\boldsymbol{y}^T(\boldsymbol{P}_{X_{2|1}}-\boldsymbol{P}_{Z_{2|1}})\boldsymbol{y}}{\hat{\sigma}^2} + \log(n)p_2= 
[\text{r}(\boldsymbol{X}_{2|1})-p_2]F_{Z_{2|1}} + \log(n)p_2\ .\
\]
Both the all-subsets analysis and the guided subsets analysis will best be able to select the true model when $F_{Z_{2|1}}$ is inflated when the model's second-order terms excludes terms in $\mathcal{A}$. Therefore, we seek a design whose expected $\boldsymbol{y}^T(\boldsymbol{P}_{X_{2|1}}-\boldsymbol{P}_{Z_{2|1}})\boldsymbol{y}$ is largest across all possible models of a given size. Note that   
\[
E[\boldsymbol{y}^T(\boldsymbol{P}_{X_{2|1}}-\boldsymbol{P}_{Z_{2|1}})\boldsymbol{y}] = [\text{r}(\boldsymbol{X}_{2|1})-p_2]\sigma^2+\boldsymbol{\beta}_2^T\boldsymbol{X}_2^T(\boldsymbol{P}_{X_{2|1}}-\boldsymbol{P}_{Z_{2|1}})\boldsymbol{X}_2\boldsymbol{\beta}_2\ ,\
\]
which involves a quadratic form of the excluded $\boldsymbol{\beta}_2$.

We initially considered taking expectation of this quantity with respect to the prior distribution of $\frac{1}{\sigma}\boldsymbol{\beta}_2$. 
 However, we are interested in a design that inflates the reduced lack-of-fit sum of squares when the excluded second-order terms involve few active effects, which does not match our prior distribution. For a fitted model involving $p_2$ second-order terms, we sum the $\text{r}(\boldsymbol{X}_{2|1})-p_2$ smallest diagonal elements of $\boldsymbol{X}_2^T(\boldsymbol{P}_{X_{2|1}}-\boldsymbol{P}_{Z_{2|1}})\boldsymbol{X}_2$, excluding the diagonal elements corresponding to the fitted second-order terms.  The intention is to assess a design's worst-case scenario by its smallest possible value for these sum of squares. Our proposed reduced lack-of-fit or rLOF criterion identifies the smallest such sum across all collections of $p_2$ second-order terms. The rLOF criterion requires specification of $p_2$ and evaluation across many subsets of second-order terms. Following our simulation study, we set $p_2=\lfloor \text{r}(\boldsymbol{X}_{2|1})/2 \rfloor$ and consider a maximum of $5,000$ models. If the total number of models exceeds $5,000$ we use a random sample of that size. The minimum value across all considered models is the final criterion value. The best design maximizes the rLOF value.

Algorithmic optimization of the rLOF criterion will be computationally expensive. We instead adopt a constrained optimization approach in which we evaluate all generated designs under \eqref{eqn:ECInobias} with $\tau^2=20$  (to emphasize alias minimization) and retain a collection of ECI-efficient designs. Then, among this collection, we choose the one that maximizes the rLOF criterion. Using the interpretation of the ECI criterion, the collection of designs are chosen based on a desired detectable signal-to-noise ratio, $S$, for the $\alpha$ considered. If a design's criterion value is less than $S$ we consider it for the constrained rLOF-criterion. We recommend $S=1$, meaning the expected variation from changing that factor from its lowest setting to highest setting equals $\sigma$. If no design's criterion value is below this threshold, we recommend re-evaluating the designs under an ECI criterion with a larger $\alpha$ or smaller $\tau^2$. 

\section{Examples and simulation study}

This section demonstrates our constrained design criterion and the benefits of the resulting optimal designs for the all-subsets analysis method of Section~4. We consider three design scenarios and include established competitors from the design literature. The first scenario is the motivating example from Section~1, with $k=5$ and $n=12$ assuming a two-factor interaction model. The last two scenarios consider a full quadratic model and settings for $k$ and $n$ that allow for some form of an ADSD as a competitor.

For each scenario, we performed our construction algorithm using $\alpha=0.05$ and $\tau^2=\{1,20\}$. If the best design's ECI value was greater than $S=1$, we increased $\alpha$ to $0.10$.  We performed the algorithm for all combinations $r \in \{0,1,2\}$ and $\ell \in \{0,1\}$ except for $(r,\ell)=(0,0)$. Setting $\ell=2$ did not noticeably improve our algorithm's ability to find a design with more lack-of-fit degrees of freedom. Each setting used $2,000$ randomly generated designs, a total of $10,000$ designs. We also retained designs generated in the middle of each iteration with an ECI criterion value less than $1$ to be sure to have a large collection of designs for optimization of the constrained rLOF criterion.

In addition to comparing the designs' structural properties, we compared the designs via a simulation study for different second-order model assumptions using our all-subsets analysis. Active effects were generated from an exponential distribution with mean $1$ with lower bound set to $2.5$, assigning positive and negative signs at random.  The active main effects were assigned at random and all active second-order terms followed the strong heredity principle. Unless otherwise stated, $\sigma^2=1$ and we performed each case $100$ times. We summarize the simulation study with the following metrics:
\begin{enumerate}
    \item $TPR_\mathcal{F}$ and $FPR_\mathcal{F}$: the proportion of times an active (inactive) factor was declared active (inactive);
    \item $\widehat{\mathcal{F}}=\mathcal{F}$\%: the proportion of times the statistically significant main effects ($\widehat{\mathcal{F}}$) are comprised of only and all factors in $\mathcal{F}$;
    \item $TPR_{2FI}$ and $FPR_{2FI}$: the proportion of times an active (inactive) two-factor interaction was declared active (inactive);
    \item $TPR_{Q}$ and $FPR_Q$: the proportion of times an active (inactive) quadratic effect was declared active (inactive).
    \item $\widehat{\mathcal{A}}=\mathcal{A}$\%: the proportion of times the final model chosen by the all-subsets analysis ($\widehat{\mathcal{A}}$) is comprised of only and all effects in $\mathcal{A}$;
    \item $|\widehat{\mathcal{A}} |$: average model size.
\end{enumerate}

\subsection{$k=5$, $n=12$, two-factor interaction model}

Recall the EDMA design and the ECI design (see Table~\ref{tab:chap2_sect1_bestdesign}) from Section~1. Both designs had 0 aliasing and, while the EDMA design had generally smaller design standard errors, it had only 1 lack-of-fit degree of freedom. The best constrained rLOF design in Table~\ref{tab:chap2_sect1_bestdesign} was consistently found for $\alpha=0.10$, except when $\ell = 1$. It has $\text{r}(\boldsymbol{X}_{2|1})=4$ and its rLOF criterion value for models of size $p_2=2$ was $7.34$.

The full factorial analysis concluded the model had  3 active main effects and 2 two-factor interactions, which we treat as the ground truth. The EDMA detected one main effect and, following the strong heredity principle, would not consider any second-order terms. We focus our attention on the all-subsets analysis for our best constrained rLOF design. First, we considered model selection for the 9 potential responses from Table~\ref{tab:chap2_sect1_bestdesign}. We then simulated $100$ datasets based on the estimated model of the full-factorial data, while randomly assigning which $3$ of the $5$ factors were important. We then repeated the simulation but added a two-factor interaction between factors $2$ and $5$ with effect size $10$, so that all interactions were included among the 3 active main effects.

Table~\ref{tab:chap2_sect5_k5n12} reports the model selection metrics for the 3 simulation scenarios. For the original reactor data, the all-subsets analysis had trouble identifying both interaction effects, which can partially be explained by the low $\widehat{\mathcal{F}}=\mathcal{F}$\%. For the comparable simulated data set, the main effect analysis improved slightly which further improved the all-subsets analysis. While  $TPR_{2FI}$ increased to $0.675$, the analysis was only identified $\mathcal{A}$ $26$\% of the time. The last simulation scenario decreased $TPR_{2FI}$ but identified $\mathcal{A}$ $40$\% of the time. This behavior has a simple explanation: when the main effect analysis perfectly recovers the important effects, the all-subsets analysis tends to perform well because it compares a smaller collection of models that are comprised primarily of active interactions. However, the analysis conditioned on this event was prone to selecting all 3 of the interaction effects. The main benefit then of the constrained rLOF design over the EDMA for this example was its ability to estimate and test main effects without bias using traditional inference. 

\begin{table}[ht]
    \centering
    \caption{Simulation study results for the best constrained rLOF design.}
    \label{tab:chap2_sect5_k5n12}
    \begin{tabular}{l|ccc|cccc}
        Scenario & $TPR_\mathcal{F}$ & $FPR_\mathcal{F}$ & $\widehat{\mathcal{F}}=\mathcal{F}$\%
        & $TPR_{2FI}$ & $FPR_{2FI}$ & $\widehat{\mathcal{A}}=\mathcal{A}$\%
        & $|\widehat{\mathcal{A}}|$
        \\ \hline 
        Reactor Data & $0.741$ & $0$ & $0.333$ & $0.611$ & $0.042$ & $0$ & $3.778$ \\
        Sim Reactor & $0.817$ & $0.110$ & $0.400$ & $0.675$ & $0.044$ & $0.260$ & $4.370$\\
        Sim Reactor 2 & $0.817$ & $0.110$ & $0.400$ & $0.635$ & $0.091$ & $0.400$ & $4.670$\\
    \end{tabular}
\end{table}

\subsection{$k=6$, $n=17$, quadratic model}

\begin{figure}[ht]
    \centering
        \includegraphics[width=0.4\textwidth]{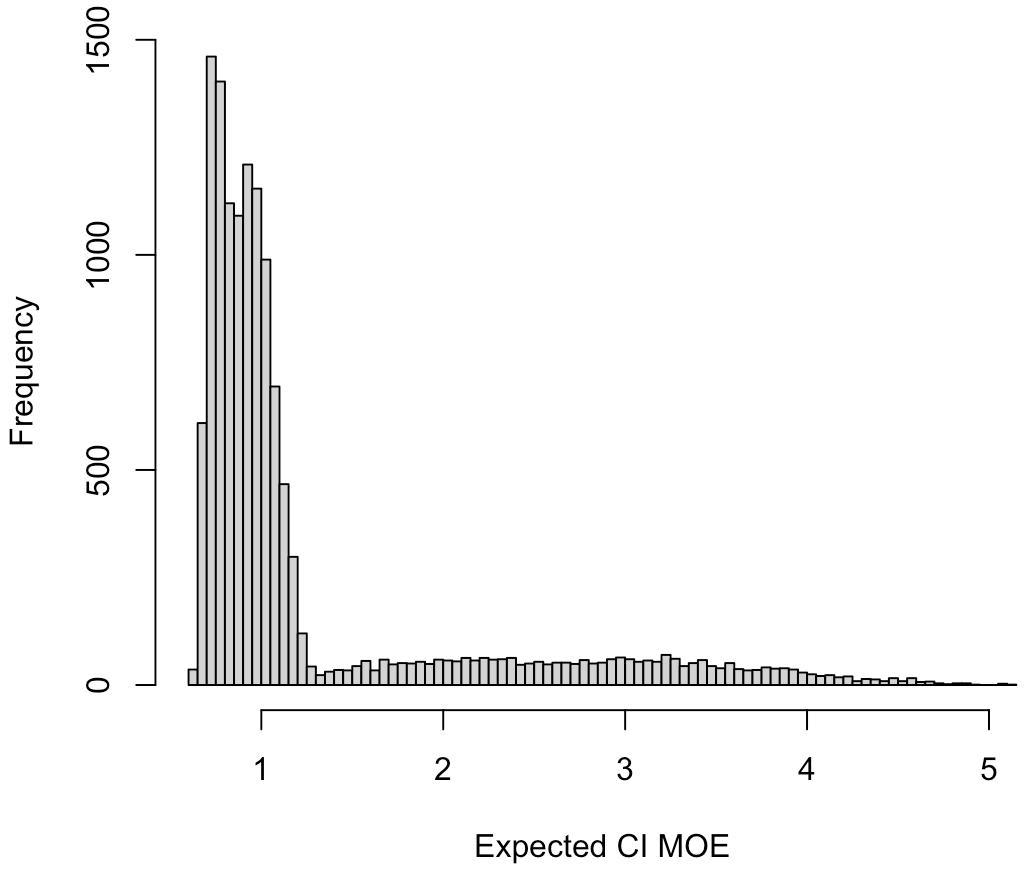}
        \includegraphics[width=0.4\textwidth]{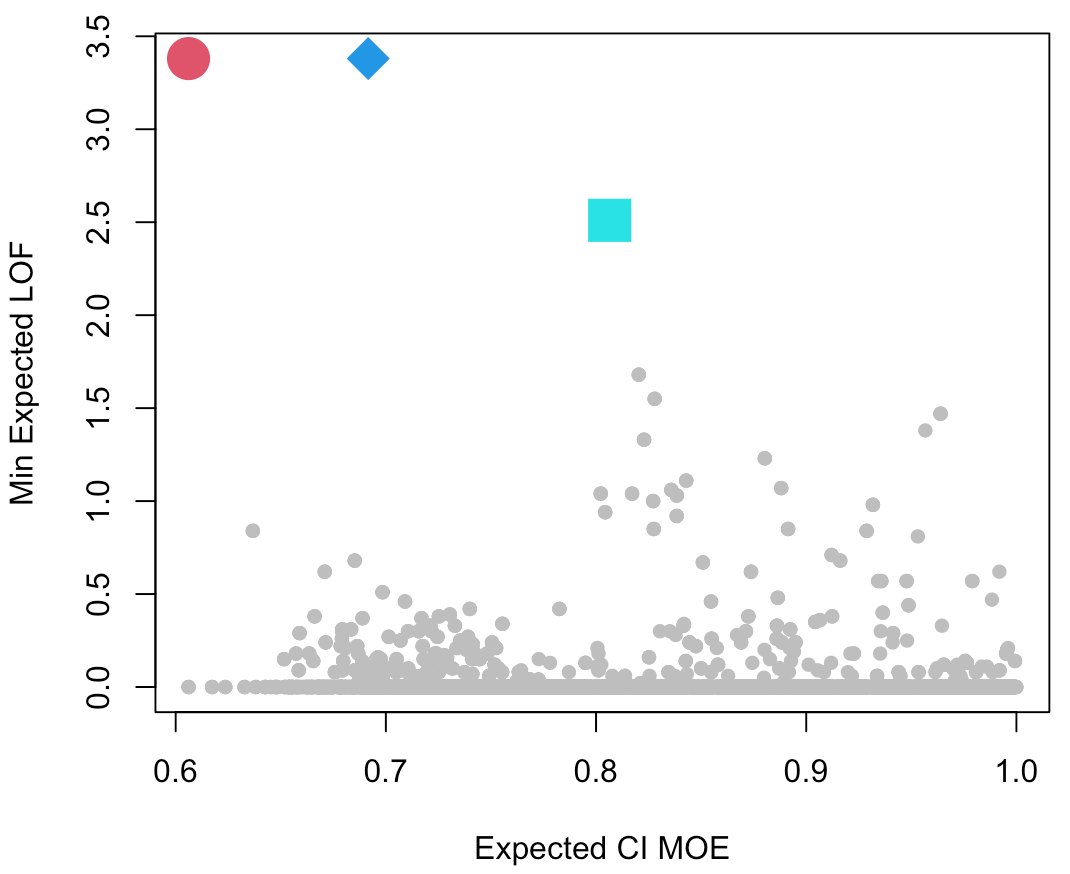}
    \caption{(Left) Histogram of ECI criterion values for all algorithmically-generated designs $(k=6,n=17)$. (Right) Scatterplot of ECI and rLOF criterion values for the subset of designs with an ECI criterion value of 1 or less. The red circle is the  utopia point (i.e. the combination of the minimum ECI value and maximum rLOF criterion value) and did not correspond to any generated design. The best design  identified by our search algorithm under the constrained rLOF criterion is shown as a light blue square. The ADSD, which was not identified by our algorithm, is shown by a dark blue diamond.}
    \label{fig:chap2_sect5_k6n17_hist_scatter}
\end{figure}

For this setting we chose $\alpha=0.10$ and, with an ADSD as our main competitor, we evaluated the constrained rLOF criterion with $p_2=4$. Combined designs gave over $8,000$ candidate designs with an ECI criterion value less than $1$. The construction algorithm identified designs with different combinations of pure errorx and rLOF degrees of freedom. Figure~\ref{fig:chap2_sect5_k6n17_hist_scatter} shows the histogram of the ECI criterion values across all generated designs as well as the scatterplot of these values and the rLOF criterion values for the constrained design set. The red circle in the upper left corner is the  utopia point (i.e. the combination of the minimum ECI value and maximum rLOF criterion value) and did not correspond to any generated design. The best design we found under the constrained rLOF criterion is shown as a light blue square. Our algorithm unfortunately did not identify the ADSD which is shown as a dark blue diamond. Clearly the ADSD would be the best constrained rLOF design had our algorithm found it.

The best constrained rLOF design we found is shown in Table~\ref{tab:chap2_sect5_k6n17_bestdesign}. Just like the ADSD, the design has $\boldsymbol{A}=0$ and $2$ rLOF degrees of freedom for estimating $\sigma^2$. While this design is Pareto dominated by the ADSD, the generated design should be fairly competitive with the ADSD. We evaluate this claim with the following simulation study.

\begin{table}[ht]
    \centering 
    \caption{Best identified constrained rLOF design for $k=6$ and $n=17$ where factors correspond to rows and the columns correspond to the individual runs. The runs are arranged as foldover pairs with an appended center run.}
    \begin{tabular}{|c|ccccccccccccccccc|}
    \hline
    1 & $+$ & $-$ & $+$ & $-$ & $-$ & $+$ & $+$ & $-$ & $0$ & $0$ & $0$ & $0$ & $-$ & $+$ & $+$ & $-$ & $0$\\ 
    2 & $-$ & $+$ & $+$ & $-$ & $0$ & $0$ & $+$ & $-$ & $+$ & $-$ & $+$ & $-$ & $+$ & $-$ & $-$ & $+$ & $0$\\
    3 & $0$ & $0$ & $+$ & $-$ & $+$ & $-$ & $-$ & $+$ & $0$ & $0$ & $-$ & $+$ & $-$ & $+$ & $-$ & $+$ & $0$\\
    4 & $0$ & $0$ & $-$ & $+$ & $+$ & $-$ & $+$ & $-$ & $+$ & $-$ & $-$ & $+$ & $-$ & $+$ & $0$ & $0$ & $0$\\
    5 & $-$ & $+$ & $0$ & $0$ & $0$ & $0$ & $+$ & $-$ & $-$ & $+$ & $+$ & $-$ & $-$ & $+$ & $0$ & $0$ & $0$\\
    6 & $+$ & $-$ & $-$ & $+$ & $0$ & $0$ & $+$ & $-$ & $-$ & $+$ & $-$ & $+$ & $0$ & $0$ & $-$ & $+$ & $0$\\
    \hline
    \end{tabular}
    \label{tab:chap2_sect5_k6n17_bestdesign}
\end{table}

Our simulation study considered $4$ combinations of active main effects, interactions, and quadratic effects shown in Table~\ref{tab:chap2_sect5_k6n17_simstudy}. The table also includes the model selection metrics for the ADSD and our best constrained rLOF design. With $2$ active main effects, the two designs performed comparably with respect to most metrics. The best constrained rLOF design performed slightly better in the case of a single quadratic effect, which can be explained by the design's slightly better estimation of quadratic effects compared to interactions. The ADSD establishes its dominance once more main effects and second-order terms are included in the model. This was expected by its larger rLOF criterion value compared to our design. Its improvement would be even more apparent if we decreased the effect sizes. Still, our generated design was competitive to the ADSD and would be a useful design in practice.


\begin{table}[ht]
    \centering
     \caption{Simulation study cases considered comparing the $k=6$, $n=17$ ADSD and the best constrained rLOF design under an all-subsets analysis.}
    \resizebox{\columnwidth}{!}{%
    \begin{tabular}{cccc|ccc|cc|cc|cc}
    \multicolumn{3}{c}{\# Active} & \multicolumn{9}{c}{}\\
    Main & 2FIs & Quad & Design & $TPR_\mathcal{F}$ & $FPR_\mathcal{F}$ & $\widehat{\mathcal{F}}=\mathcal{F}$\%
        & $TPR_{2FI}$ & $FPR_{2FI}$ & $TPR_{Q}$ & $FPR_{Q}$ & $\widehat{\mathcal{A}}=\mathcal{A}$\%
        & $|\widehat{\mathcal{A}}|$\\ \hline
    2 & 1 & 0 & ADSD & $1.000$ & $0.088$ & $0.770$ & $1.000$ & $0.021$ & $0$ & $0.115$ & $0.460$ & $4.330$\\
      &   &   & Best rLOF & $1.000$ & $0.100$ & $0.820$  & $0.98$ & $0.029$ & $0$ & $0.096$ & $0.500$ & $4.370$ \\ \hline
    2 & 0 & 1 & ADSD & $1.000$ & $0.088$ & $0.770$ & $0$ & $0.020$ & $0.980$ & $0.086$ & $0.580$ & $5.050$\\
      &   &   & Best rLOF & $1.000$ & $0.100$ & $0.820$  & $0$ & $0.023$ & $0.970$ & $0.064$ & $0.690$ & $5.030$ \\ \hline
    3 & 2 & 2 & ADSD & $1.000$ & $0.082$ & $0.820$ & $0.975$ & $0.035$ & $0.935$ & $0.090$ & $0.530$ & $7.880$\\
      &   &   & Best rLOF & $1.000$ & $0.093$ & $0.840$ & $0.940$ & $0.045$ & $0.885$ & $0.063$ & $0.510$ & $7.760$ \\ \hline
    4 & 4 & 3 & ADSD & $1.000$ & $0.055$ & $0.900$ & $0.655$ & $0.093$ & $0.590$ & $0.200$ & $0.110$ & $10.120$\\
      &   &   & Best rLOF & $1.000$ & $0.115$ & $0.820$ & $0.608$ & $0.102$ & $0.510$ & $0.210$ & $0.010$ & $9.940$ \\ \hline
    \end{tabular}
    }
    \label{tab:chap2_sect5_k6n17_simstudy}
\end{table}


\subsection{$k=7$, $n=24$, quadratic model}

\cite{leonard2017bayesian} constructed a 3-level screening design with $k=7$ and $n=24$ having 6 pure error degrees of freedom. Their design, however, has biased main effects for the quadratic model with an average and maximum absolute aliasing value of $0.147$ and $0.593$, respectively. The design's ECI value with $\alpha=0.05$ and $\tau^2=1$ is $1.415$. These properties tell us this design is not a strong competitor and so will not be considered further. A stronger competing design can be constructed using the ADSD approach with $f=5$ fake factors and removing the center run. Removal of the center run maintains the design's $5$ lack-of-fit degrees of freedom, zero aliasing, and small design standard errors for the main effect model.

\begin{figure}[ht]
    \centering
    \includegraphics[width=0.5\textwidth]{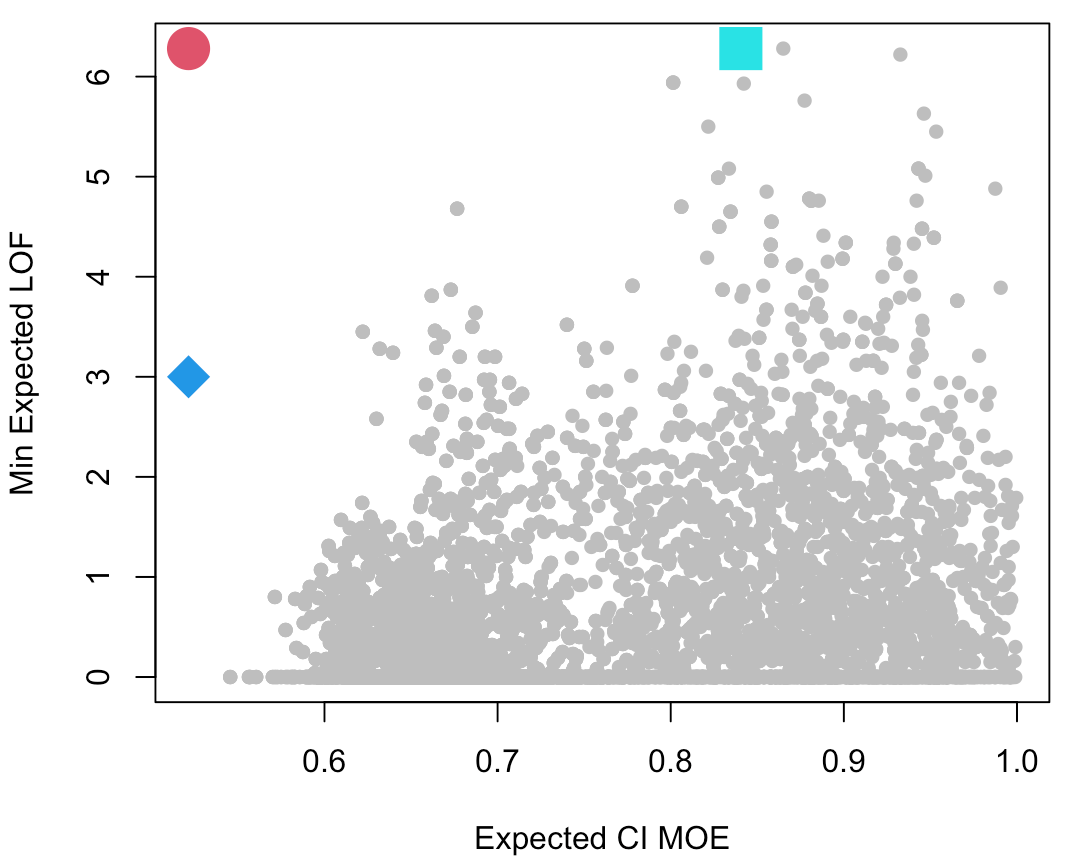}
    \caption{Scatterplot of ECI and rLOF criterion values for the subset of designs $(k=7,n=24)$ with an ECI criterion value of 1 or less. The red circle is the  utopia point and did not correspond to any generated design. The best design under the constrained rLOF criterion is shown as a light blue square. The ADSD, which was not identified by our algorithm, is shown by a dark blue diamond.}
    \label{fig:chap2_sect5_k7n24_scatter}
\end{figure}

When constructing designs under the constrained rLOF criterion, we found the generated designs were often similar regardless of the choice of $r$ and $\ell$. The constrained set of designs had at least $4$ degrees of freedom to estimate $\sigma^2$, mostly coming from lack-of-fit (just like the ADSD) rather than pure error (the approach by \cite{leonard2017bayesian}). Figure~\ref{fig:chap2_sect5_k7n24_scatter} shows the scatterplot of the ECI and rLOF criterion values for the collection of designs with the same coloring of points as in Figure~\ref{fig:chap2_sect5_k6n17_hist_scatter}.

The algorithm again failed to identify the proposed ADSD, which optimized the ECI criterion. However, the best constrained rLOF design is expected to have better analysis properties in the all-subsets analysis. This design, shown in Table~\ref{tab:chap2_sect5_k7n24_bestdesign}, has 4 lack-of-fit degrees of freedom and negligible but nonzero aliasing. The average absolute aliasing value is $0.004$ and the maximum value is $0.034$. Note that aliasing will most significantly impact the first stage of the analysis, although clearly issues with the first stage analysis will impact the all-subsets regression.

\begin{table}[ht]
    \centering 
    \caption{Best identified constrained rLOF design for $k=7$ and $n=24$. For ease of presentation, we show $\boldsymbol{D}^T$, where factors correspond to rows and the columns correspond to the individual runs. The first 22 runs are arranged as foldover pairs.}
    \resizebox{\columnwidth}{!}{%
    \begin{tabular}{|c|cccccccccccccccccccccccc|}
    \hline
    1 &  $+$ &	$-$	& $+$ &	$-$ & $+$ &	$-$	& $+$ &	$-$	& $-$ & $+$	& $+$	& $-$	& $+$ &	$-$ & $0$ &	 $0$ &	$+$ & $-$ &	$+$ & $-$ &	$-$ &	$+$	& $0$ &	$0$ \\
    2 &  $-$ &  $+$	& $-$ &	$+$	& $0$ &	$0$	& $+$ &	$-$	& $0$ &	$0$	& $+$	& $-$	& $+$ &	$-$ & $-$ &	 $+$ &	$-$	& $+$ &	$-$ & $+$ &	$+$ &	$-$	& $+$ &	$0$ \\
    3 &  $0$ &	$0$	& $0$ &	$0$	& $0$ &	$0$	& $+$ &	$-$	& $+$ &	$-$	& $-$	& $+$	& $0$ &	$0$ & $0$ &	 $0$ &	$-$	& $+$ &	$-$ & $+$ &	$-$ &	$+$ &	 $0$ &	$0$ \\
    4 &  $-$ &	$+$	& $+$ &	$-$	& $-$ &	$+$	& $0$ &	$0$	& $+$ &	$-$	& $0$	& $0$	& $-$ &	$+$	& $-$ &	 $+$ &	$+$	& $-$ &	$+$ & $-$	& $0$ &	$0$	& $-$ &	$0$ \\
    5 &  $+$ &	$-$	& $-$ &	$+$	& $+$ &	$-$	& $+$ & $-$	& $+$ &	$-$	& $0$	& $0$	& $-$ &	$+$	& $+$ &	 $-$ &	$+$	& $-$ &	$0$ & $0$	& $+$ &	$-$	& $0$ &	$0$ \\
    6 &  $0$ &	$0$	& $+$ &	$-$	& $+$ &	$-$	& $-$ &	$+$	& $0$ &	$0$	& $-$	& $+$	& $+$ &	$-$	& $-$ &	 $+$ &	$-$	& $+$ &	$+$ & $-$	& $+$ &	$-$	& $0$ &	$0$ \\
    7 &  $-$ &	$+$	& $+$ &	$-$	& $+$ &	$-$	& $-$ &	$+$	& $+$ &	$-$	& $+$	& $-$	& $-$ &	$+$	& $0$ &	 $0$ &	$+$	& $-$ &	$+$ & $-$	& $-$ &	$+$ & $-$ &	$0$ \\
    \hline
    \end{tabular}
    }
    \label{tab:chap2_sect5_k7n24_bestdesign}
\end{table}

A simulation protocol like in Section~5.2 was performed for this design and the ADSD; the model selection results are given in Table~\ref{tab:chap2_sect5_k7n24_simstudy}. Given the smaller ECI values for this situation, the off-set for main effects was reduced from $2.5$ to $1.5$. With $3$ active main effects, the two designs performed comparably with respect to most metrics. The best constrained rLOF design performed slightly better for detecting quadratic effects except for the most saturated model. Note the $TPR_\mathcal{F}$ decreases for the best constrained rLOF design, since the aliasing increases slightly. The constrained rLOF design is mostly dominant for the scenario of $5$ active main effects, $4$ two-factor interactions, and $3$ quadratic effects. The design is superior for all second-order term metrics, and it was able to perfectly recover the correct model $21$\% of the time. Again, this was expected by its larger rLOF criterion value compared to the ADSD.


\begin{table}[ht]
    \centering
    \caption{Simulation study cases considered comparing the $k=7$, $n=24$ ADSD and the best constrained rLOF design under an all-subsets analysis.}
    \resizebox{\columnwidth}{!}{%
    \begin{tabular}{cccc|ccc|cc|cc|cc}
    \multicolumn{3}{c}{\# Active} & \multicolumn{9}{c}{}\\
    Main & 2FIs & Quad & Design & $TPR_\mathcal{F}$ & $FPR_\mathcal{F}$ & $\widehat{\mathcal{F}}=\mathcal{F}$\%
        & $TPR_{2FI}$ & $FPR_{2FI}$ & $TPR_{Q}$ & $FPR_{Q}$ & $\widehat{\mathcal{A}}=\mathcal{A}$\%
        & $|\widehat{\mathcal{A}}|$\\ \hline
    3 & 1 & 1 & ADSD & $1.000$ & $0.038$ & $0.880$ & $1.000$ & $0.026$ & $0.980$ & $0.068$ & $0.450$ & $6.050$\\
      &   &   & Best rLOF & $1.000$ & $0.033$ & $0.900$  & $1.000$ & $0.021$ & $0.990$ & $0.078$ & $0.450$ & $6.010$ \\ \hline
    3 & 2 & 2 & ADSD & $1.000$ & $0.038$ & $0.880$ & $0.995$ & $0.019$ & $0.955$ & $0.044$ & $0.570$ & $7.640$\\
      &   &   & Best rLOF & $0.997$ & $0.028$ & $0.910$  & $0.995$ & $0.011$ & $0.980$ & $0.044$ & $0.670$ & $7.470$ \\ \hline
    3 & 3 & 3 & ADSD & $1.000$ & $0.038$ & $0.880$ & $1.000$ & $0.010$ & $0.953$ & $0.010$ & $0.820$ & $9.230$\\
      &   &   & Best rLOF & $0.987$ & $0.028$ & $0.870$  & $0.947$ & $0.004$ & $0.947$ & $0.005$ & $0.790$ & $8.850$ \\ \hline
    5 & 3 & 1 & ADSD & $1.000$ & $0.065$ & $0.890$ & $0.983$ & $0.053$ & $0.890$ & $0.077$ & $0.310$ & $10.39$\\
      &   &   & Best rLOF & $0.994$ & $0.045$ & $0.910$ & $0.940$ & $0.055$ & $0.920$ & $0.100$ & $0.320$ & $10.39$ \\  \hline
     5 & 4 & 3 & ADSD & $1.000$ & $0.065$ & $0.890$ & $0.753$ & $0.109$ & $0.537$ & $0.155$ & $0.130$ & $12.23$\\
      &   &   & Best rLOF & $0.994$ & $0.050$ & $0.880$ & $0.760$ & $0.102$ & $0.700$ & $0.125$ & $0.210$ & $12.45$ \\ \hline
    \end{tabular}
    }
    
    \label{tab:chap2_sect5_k7n24_simstudy}
\end{table}

\section{Discussion}\label{s:Discussion}

This paper presents a new constrained optimization criterion targeting a two-stage analysis of a screening design that requires unbiased estimation of $\sigma^2$, either through pure error or lack-of-fit degrees of freedom. The ECI criterion naturally balances the relative importance of aliasing, design variance, and error degrees of freedom for the main effect model analysis. To construct optimal designs under this criterion, we developed a coordinate exchange algorithm for fast updates when exchanging coordinates of replicated rows simultaneously. The constrained rLOF criterion conditions on a large set of ECI-efficient designs and was motivated by the proposed all-subsets BIC selection method that was shown to generally have superior selection properties compare to the guided subsets approach from \cite{jones2017effective}. The designs generated in this paper either beat established screening designs, or would have declared such designs optimal under the constrained rLOF criterion.

When using the constrained rLOF criterion, practitioners should recognize that not every $n$ and $k$ combination will yield screening designs that balance all three desirable design properties for the ECI criterion. This is mainly due to the limited selection of designs for some $n$ and $k$ combinations. In this case we recommend finding a design with two degrees of freedom for estimation $\sigma^2$ and try one of two strategies:
\begin{enumerate}
    \item Identify a design with as small aliasing as possible so that a successful first stage analysis may be performed
    \item Try to balance minimizing both aliasing and design variance and replace the two-stage analysis with an overall all-subsets analysis.
\end{enumerate}
The latter analysis strategy will lead to a more computationally heavily analysis due to the inclusion of $k$ additional factors in the all-subsets analysis. 

For quadratic models, we often found that designs with improved ability to test for quadratic effects had to sacrifice some testing power for interaction effects.  This is because the minimum possible variance for a quadratic effect tends to be larger than that for main effects and interactions. This behavior was recently noted by \cite{allen2021incorporating} who proposed incorporating minimum variances in design criteria.

There are many areas of future work that we hope to pursue. First, 
while our algorithmic design construction approach often found better or comparable designs to existing competitors, the algorithm can be slow for larger designs. It is important then to identify construction techniques like the ADSD to identify designs that minimize aliasing and have degrees of freedom to estimate error. Another area worth pursuing is to address the tendency of the all-subsets analysis to overselect second-order terms. For some problems, the mBIC penalty does not grow fast enough to discount overfitting. It would be worth considering a Bayesian model averaging approach like in \cite{marley2010comparison}.






\newpage
\section{Supplementary Materials}

\subsection{Designs for Reactor Experiment}

\begin{table}[ht]
    \centering
    \begin{tabular}{|cccccc|cccccc|cccccc|}
     \multicolumn{6}{c}{(a) NRFFD} & \multicolumn{6}{c}{(b) Bayesian $D$} & \multicolumn{6}{c}{(c) EDMA}  \\
    \hline
    1 & 2 & 3 & 4 & 5 & $y$ &
    1 & 2 & 3 & 4 & 5 & $y$ &
    1 & 2 & 3 & 4 & 5 & $y$ \\
    \hline
    $-$ & $+$ & $-$ & $+$ & $+$ & 78 &
    $-$ & $-$ & $+$ & $+$ & $-$ & 66 & 
    $+$ & $-$ & $-$ & $+$ & $-$ & 61 \\
    $-$ & $+$ & $+$ & $+$ & $-$ & 95 &
    $-$ & $-$ & $-$ & $-$ & $-$ & 61 &
    $-$ & $+$ & $+$ & $-$ & $+$ & 67 \\
    $-$ & $-$ & $+$ & $-$ & $+$ & 59 &
    $+$ & $-$ & $-$ & $+$ & $-$ & 61 &
    $+$ & $+$ & $+$ & $+$ & $-$ & 98 \\
    $+$ & $+$ & $+$ & $+$ & $+$ & 82 & 
    $-$ & $+$ & $-$ & $-$ & $+$ & 70 & 
    $-$ & $-$ & $-$ & $-$ & $+$ & 56 \\
    $+$ & $+$ & $+$ & $-$ & $-$ & 61 &
    $+$ & $-$ & $-$ & $-$ & $+$ & 63 & 
    $+$ & $+$ & $+$ & $-$ & $-$ & 61 \\
    $+$ & $-$ & $-$ & $+$ & $-$ & 61 &
    $+$ & $-$ & $+$ & $+$ & $+$ & 42 & 
    $-$ & $-$ & $-$ & $+$ & $+$ & 44 \\
    $-$ & $+$ & $-$ & $-$ & $+$ & 70 &
    $-$ & $+$ & $-$ & $+$ & $-$ & 94 &
    $+$ & $-$ & $-$ & $-$ & $+$ & 63 \\
    $+$ & $-$ & $+$ & $+$ & $+$ & 42 &
    $-$ & $-$ & $-$ & $+$ & $+$ & 44 &
    $-$ & $+$ & $+$ & $+$ & $-$ & 95 \\
    $+$ & $+$ & $-$ & $-$ & $-$ & 61 &
    $-$ & $-$ & $+$ & $-$ & $+$ & 59 &
    $+$ & $+$ & $-$ & $+$ & $+$ & 77 \\
    $-$ & $-$ & $+$ & $-$ & $-$ & 53 &
    $-$ & $+$ & $+$ & $+$ & $+$ & 81 &
    $-$ & $-$ & $+$ & $-$ & $-$ & 53 \\
    $+$ & $-$ & $-$ & $-$ & $+$ & 63 &
    $+$ & $+$ & $+$ & $-$ & $-$ & 61 &
    $-$ & $+$ & $-$ & $-$ & $-$ & 63 \\
    $-$ & $-$ & $-$ & $+$ & $-$ & 69 &
    $+$ & $+$ & $-$ & $+$ & $+$ & 77 &
    $+$ & $-$ & $+$ & $+$ & $+$ & 42 \\ \hline
    \end{tabular}
    \caption{Potential fractional factorial designs with $n=12$ runs and $5$ two-level factors. The design settings are represented by $\pm$ instead of $\pm 1$. The responses are taken from the corresponding rows of the $2^5$ full-factorial Reactor experiment in \cite{box1978statistics}.}
    \label{tab:chap2_sect1_3designs}
\end{table}

\subsection{Maximum deviation of ECI from $\beta_j/\sigma$}

For a confidence interval under the first stage analysis to achieve the desired $100(1-\alpha)$ confidence level with the smallest possible margin of error, we need the ECI to be centered at $\beta_j/\sigma$ and for $c(\alpha,g)\sqrt{v_j}$ (as defined in the main article) to be small. Both of these qualities may be measured by calculating the largest distance of a value in the ECI from $\beta_j/\sigma$, which will correspond to one of the two endpoints of the interval:
\begin{align*}
&\max\left( \ \left|(\beta_j + \boldsymbol{A}_j\boldsymbol{\beta}_2)/\sigma - c(\alpha,g) \sqrt{v_j} -\beta_j/\sigma\right|, \  \left|(\beta_j + \boldsymbol{A}_j\boldsymbol{\beta}_2)/\sigma + c(\alpha,g)\sqrt{v_j} -\beta_j/\sigma \right| \ \right)\\ 
\Leftrightarrow & \max\left( \ \left| \boldsymbol{A}_j\boldsymbol{\beta}_2/\sigma - c(\alpha,g) \sqrt{v_j} \right|, \  \left|\boldsymbol{A}_j\boldsymbol{\beta}_2/\sigma + c(\alpha,g)\sqrt{v_j}\right| \ \right)\ .\ 
\end{align*}
As $c(\alpha,g)\sqrt{v_j} > 0$, when $\boldsymbol{A}_j\boldsymbol{\beta}_2<0$, $\boldsymbol{A}_j\boldsymbol{\beta}_2=-|\boldsymbol{A}_j\boldsymbol{\beta}_2|$ and the maximum is $| \boldsymbol{A}_j\boldsymbol{\beta}_2/\sigma - c(\alpha,g) \sqrt{v_j}|$, which may be written as $\frac{1}{\sigma}|\boldsymbol{A}_j\boldsymbol{\beta}_2|+c(\alpha,g) \sqrt{v_j}$. Otherwise, the maximum is $\left|\boldsymbol{A}_j\boldsymbol{\beta}_2/\sigma + c(\alpha,g)\sqrt{v_j}\right|$ which again may be written as $\frac{1}{\sigma}|\boldsymbol{A}_j\boldsymbol{\beta}_2|+c(\alpha,g) \sqrt{v_j}$ since $\boldsymbol{A}_j\boldsymbol{\beta}_2 = |\boldsymbol{A}_j\boldsymbol{\beta}_2|$. Therefore, the maximum deviation of the interval from $\beta_j/\sigma$ is 
\[
\frac{1}{\sigma}|\boldsymbol{A}_j\boldsymbol{\beta}_2|+c(\alpha,g) \sqrt{v_j}
\]
as claimed in the main article.

\subsection{Expected value of $\sqrt{\hat{\sigma}_X^2}$} \label{chapter2technical1}
If $\boldsymbol{e}\sim N(0,\sigma^2\boldsymbol{I})$ then $Z=g \frac{\hat{\sigma}^2}{\sigma^2}$ follows a $\chi^2$ distribution with $g$ degrees of freedom. We want to derive $E(\sqrt{\hat{\sigma}^2})=\frac{\sigma}{\sqrt{g}} E(\sqrt{Z})$. With $\Gamma(g)=\int_0^\infty t^{g-1}e^{-t} dt$,
\begin{align*}
    E(\sqrt{Z})&=\int_0^\infty \sqrt{z} z^{g/2 -1} \frac{e^{-x/2}}{2^{g/2}\Gamma(g/2)} dz\\
    &=\frac{2\sqrt{2}}{\Gamma(g/2)}\int_0^\infty \sqrt{\frac{z}{2}} \left(\frac{z}{2}\right)^{g/2 -1}e^{-z/2} dz\\
    &=\frac{2\sqrt{2}}{\Gamma(g/2)}\int_0^\infty \left(\frac{z}{2}\right)^{(g+1)/2 -1}e^{-z/2} dz\\
    &=\frac{\sqrt{2}}{\Gamma(g/2)}\int_0^\infty t^{(g+1)/2 -1}e^{-t} dt \qquad (t=z/2 \quad dz=2 dt)\\
    &=\sqrt{2}\frac{\Gamma\left(\frac{g+1}{2}\right)}{\Gamma(\frac{g}{2})}\ .\
\end{align*}
Hence $\frac{1}{\sigma}E(\sqrt{\hat{\sigma}^2})=\sqrt{2/g} \Gamma([g+1]/2)/\Gamma(g/2)$.
\subsection{Expected value of $\frac{1}{\sigma}|\boldsymbol{A}_j\boldsymbol{\beta}_2|$}

Assuming $\frac{1}{\sigma}\boldsymbol{\beta}_2 \sim N(0,\tau^2 \boldsymbol{I})$,  it follows that $\frac{1}{\sigma} \boldsymbol{A}_j\boldsymbol{\beta}_2$ is Normally distributed with mean $0$ and variance $\tau^2 \boldsymbol{A}_j\boldsymbol{A}_j^T$. Hence $\frac{1}{\sigma}| \boldsymbol{A}_j\boldsymbol{\beta}_2|$ follows a half-normal distribution with
\[
E\left(\frac{1}{\sigma}|\boldsymbol{A}_j\boldsymbol{\beta}_2|\right)=\sqrt{\frac{2}{\pi}}\sqrt{\text{Var}\left(\frac{1}{\sigma} \boldsymbol{A}_j\boldsymbol{\beta}_2\right)}=\sqrt{\frac{2\tau^2}{\pi}\boldsymbol{A}_j\boldsymbol{A}_j^T}\ .\
\]

\subsection{$\text{E}\left(\frac{\hat{\sigma}^2_{L}}{\sigma^2}\right)=1 + \frac{\tau^2}{\ell}\text{tr}(\boldsymbol{X}_2^T\boldsymbol{P}_{L}\boldsymbol{X}_2)$}

Recall $\hat{\sigma}^2_{L}=\boldsymbol{y}^T\boldsymbol{P}_L\boldsymbol{y}/\ell$. Using the expected value of a quadratic form, conditional on $\boldsymbol{\beta}_2$, it follows
\[
\text{E}(\hat{\sigma}^2_{L}/\sigma^2 \, | \, \boldsymbol{\beta}_2)=\frac{\sigma^2+(\boldsymbol{\beta}^T\boldsymbol{X}^T\boldsymbol{P}_L\boldsymbol{X}\boldsymbol{\beta})/\ell}{\sigma^2}=1+\frac{1}{\ell}(\boldsymbol{\beta}_2/\sigma)^T\boldsymbol{X}_2^T\boldsymbol{P}_L\boldsymbol{X}_2(\boldsymbol{\beta}_2/\sigma)\ ,\
\]
where the second equality follows from the requirement $\boldsymbol{X}_1^T\boldsymbol{L}=0$. We then take expectation of this quantity, which involves a quadratic form with respect to $\boldsymbol{\beta}_2/\sigma$, assuming $\boldsymbol{\beta}_2/\sigma$ follows a Normal distribution with mean $\boldsymbol{0}$ and variance $\tau^2\boldsymbol{I}$:
\[
\text{E}(\hat{\sigma}^2_{L}/\sigma^2)= 1 + \frac{\tau^2}{\ell}\text{tr}(\boldsymbol{X}_2^T\boldsymbol{P}_L\boldsymbol{X}_2)\ .\
\]

\subsection{$c(\alpha,g,\tau)=t_{\alpha/2,g}\sqrt{1+\frac{\tau^2}{g}\sum_{m=1}^{\ell^*} \lambda_m}$}

The goal of this section is to generalize $c(\alpha,g)=\frac{1}{\sigma}E(\sqrt{\hat{\sigma}_X^2})t_{\alpha/2,g}$ where $g=\text{r}(\boldsymbol{I}-\boldsymbol{P}_X)$, 
\[
\hat{\sigma}^2_X=\frac{SS_{PE}+SS_{LOF}}{g}=\frac{r\hat{\sigma}^2_{PE}+\ell\hat{\sigma}^2_L}{g}\ ,\
\]
and $\hat{\sigma}^2_{PE}=SS_{PE}/r$ is the pure-error estimator assuming replication. Deriving $E(\sqrt{\hat{\sigma}_X^2})$ is challenging when $\hat{\sigma}^2_L$ is biased so we use its upper bound $\sqrt{E(\hat{\sigma}^2_X)}$ and define $c(\alpha,g,\tau)=t_{\alpha/2,g}\frac{\sqrt{E(\hat{\sigma}^2_X)}}{\sigma}=t_{\alpha/2,g}\sqrt{E\left(\frac{\hat{\sigma}^2_X}{\sigma^2}\right)}.$ Using the above expression for $\hat{\sigma}^2_X$ and our construction of $\boldsymbol{L}^*$,
\begin{align*}
E\left(\frac{\hat{\sigma}^2_X}{\sigma^2}\right)=\frac{r}{g}E\left(\frac{\hat{\sigma}^2_{PE}}{\sigma^2}\right)+\frac{\ell}{g}E\left(\frac{\hat{\sigma}^2_{L}}{\sigma^2}\right)=\frac{r}{g}+\frac{\ell}{g}\left(1+\frac{\tau^2}{\ell}\sum_{m=1}^{\ell^*} \lambda_m\right)&=\frac{r+\ell}{g}+\frac{\tau^2}{g}\sum_{m=1}^{\ell^*} \lambda_m    \\
&=1+\frac{\tau^2}{g}\sum_{m=1}^{\ell^*} \lambda_m \ ,\
\end{align*}
as $E(\hat{\sigma}^2_{PE})=\sigma^2$ and $g=r+\ell$. This gives the desired result.

\subsection{Generalized Guided Subsets Approach}

The first step of the guided subsets selection procedure from \cite{jones2017effective} intends to assess the overall significance of the second order terms, after adjusting for the main effect model and $f$ fake factors assumed to be orthogonal to the main effects and all second-order terms of the main effects. The sum-of-squares from the fake factor columns, $\boldsymbol{F}$, are the lack-of-fit sum-of-squares and so correspond to the matrix $\boldsymbol{L}$ in our main article. Note $\boldsymbol{y}^T\boldsymbol{P}_{X_{2|1}}\boldsymbol{y}$, where $\boldsymbol{X}_{2|1}=(\boldsymbol{I}-\boldsymbol{P}_{X_1})\boldsymbol{X}_2$, is the general expression for the sum-of-squares for the second-order terms, after adjusting for the main effect model. No adjustment is needed for pure error or lack-of-fit because their sum-of-squares are orthogonal to those for the intercept, main effects, and all second-order terms. \cite{jones2017effective} employ the test statistic $F=(\text{TSS}/(k+f))/\hat{\sigma}^2_X$ where they define $\text{TSS}=\boldsymbol{y}_{\text{2nd}}^T\boldsymbol{y}_{\text{2nd}}$ with $\boldsymbol{y}_{\text{2nd}}=(\boldsymbol{I}-\boldsymbol{X}_{\text{DF}}(\boldsymbol{X}_{\text{DF}}^T\boldsymbol{X}_{\text{DF}})^{-1}\boldsymbol{X}_{\text{DF}}^T)\boldsymbol{y}_c$ and $\boldsymbol{y}_c$ is the centered response and $\boldsymbol{X}_{\text{DF}}$ is the matrix comprised of the centered columns of the main effects and fake factors. We will show that under the stated assumptions in \cite{jones2017effective}, $\boldsymbol{y}_{\text{2nd}}^T\boldsymbol{y}_{\text{2nd}}$ is equivalent to $\boldsymbol{y}^T\boldsymbol{P}_{X_{2|1}}\boldsymbol{y}$.

The derivations in \cite{jones2017effective} presume that the uncorrected sum of squares can be broken up into:
\[
\boldsymbol{y}^T\boldsymbol{y} = \boldsymbol{y}^T\boldsymbol{P}_1\boldsymbol{y}+\boldsymbol{y}^T\boldsymbol{P}_D\boldsymbol{y}+\boldsymbol{y}^T\boldsymbol{P}_F\boldsymbol{y}+\boldsymbol{y}^T\boldsymbol{P}_{X_2}\boldsymbol{y}
\]
where $\boldsymbol{P}_1=\frac{1}{n}\boldsymbol{1}$ and they denote $\boldsymbol{P}_{X_2}$ as the projection matrix onto the $(\boldsymbol{I}-\boldsymbol{P}_1)\boldsymbol{X}_2$, the centered matrix of second-order terms of the $k$ main effects. The four projection matrices are assumed to be mutually orthogonal and their sum equals $\boldsymbol{I}$. This implies the columns of the four vectors/matrices $\boldsymbol{1}$, $\boldsymbol{D}$, $\boldsymbol{F}$, and $\boldsymbol{X}_2$ are mutually orthogonal and $\text{r}(\boldsymbol{X}_2)=n-1-k-f=k+f$ since their DSD construction has $n=2(k+f)+1$. Additionally, our defined $\boldsymbol{P}_{X_1}=\boldsymbol{P}_{1}+\boldsymbol{P}_{D}$ and so $\boldsymbol{X}_{2|1}=(\boldsymbol{I}-\boldsymbol{P}_1-\boldsymbol{P}_D)\boldsymbol{X}_2=(\boldsymbol{I}-\boldsymbol{P}_1)\boldsymbol{X}_2$, hence their $\boldsymbol{y}^T\boldsymbol{P}_{X_2}\boldsymbol{y}=\boldsymbol{y}^T\boldsymbol{P}_{X_{2|1}}\boldsymbol{y}$

A vector or matrix is centered by premultiplying by $\boldsymbol{I}-\boldsymbol{P}_1$, making $\boldsymbol{y}_c=(\boldsymbol{I}-\boldsymbol{P}_1)\boldsymbol{y}$ and  $(\boldsymbol{I}-\boldsymbol{P}_1)\boldsymbol{X}_{\text{DF}}=\boldsymbol{X}_{\text{DF}}$. Then
\begin{align*}
\boldsymbol{y}_{\text{2nd}}&=(\boldsymbol{I}-\boldsymbol{X}_{\text{DF}}(\boldsymbol{X}_{\text{DF}}^T\boldsymbol{X}_{\text{DF}})^{-1}\boldsymbol{X}_{\text{DF}}^T)(\boldsymbol{I}-\boldsymbol{P}_1)\boldsymbol{y}\\
&=(\boldsymbol{I}-\boldsymbol{P}_1-\boldsymbol{X}_{\text{DF}}(\boldsymbol{X}_{\text{DF}}^T\boldsymbol{X}_{\text{DF}})^{-1}\boldsymbol{X}_{\text{DF}}^T)\boldsymbol{y}\\
&=(\boldsymbol{I}-\boldsymbol{P}_1-\boldsymbol{P}_D-\boldsymbol{P}_F)\boldsymbol{y}\ ,\
\end{align*}
as the above assumptions about orthogonality imply $\boldsymbol{P}_{\text{DF}}=\boldsymbol{X}_{\text{DF}}(\boldsymbol{X}_{\text{DF}}^T\boldsymbol{X}_{\text{DF}})^{-1}\boldsymbol{X}_{\text{DF}}=\boldsymbol{P}_D+\boldsymbol{P}_F$. Moreover, $\boldsymbol{I}-\boldsymbol{P}_1-\boldsymbol{P}_D-\boldsymbol{P}_F=\boldsymbol{P}_{X_2}=\boldsymbol{P}_{X_{2|1}}$ so $\boldsymbol{y}_{\text{2nd}}^T\boldsymbol{y}_{\text{2nd}}=\boldsymbol{y}^T\boldsymbol{P}_{X_{2|1}}\boldsymbol{y}$, as we claimed.

There are two additional comments for why our method is more generalized:
\begin{enumerate}
    \item The first step of the guided subsets approach requires orthogonality between the main effects and fake factors. Otherwise, their lack-of-fit estimator is incorrect. Therefore, the procedure only works when $k+f$ is even. Our formulas for calculating the lack-of-fit estimator and the sum-of-squares $\boldsymbol{y}^T\boldsymbol{P}_{X_{2|1}}\boldsymbol{y}$ do not need this assumption.
    \item Their $F$ test is incorrect in the presence of $n_c>1$ center runs, which they claim their approach can handle. The issue is that the above decomposition for $\boldsymbol{y}^T\boldsymbol{y}$ does not include the resulting pure error sum-of-squares. This causes $\boldsymbol{y}_{\text{2nd}}^T\boldsymbol{y}_{\text{2nd}}$ to include the pure error sum-of-squares.
\end{enumerate}
We also need to show that the residual sum-of-squares in the second part of the guided subsets approach equals our claimed formula: $\boldsymbol{y}^T(\boldsymbol{X}_{2|1}-\boldsymbol{Z}_{2|1})\boldsymbol{y}$. Using the previously stated orthogonality properties, the residual sum-of-squares after fitting the $p_2$ second-order terms in $\boldsymbol{Z}_{2|1}$
admits the expression
\[
\boldsymbol{y}^T(\boldsymbol{I}-\boldsymbol{P}_1-\boldsymbol{P}_D-\boldsymbol{P}_F-\boldsymbol{P}_{Z_{2|1}})\boldsymbol{y}\ .\
\]
The identity $\boldsymbol{I}=\boldsymbol{P}_1+\boldsymbol{P}_D+\boldsymbol{P}_F+\boldsymbol{P}_{X_{2|1}}$ gives the desired result.


\subsection{Equivalent mBIC Representation}

Recall the mBIC expression:
\[
    \text{mBIC}=\frac{\boldsymbol{y}^T(\boldsymbol{I}-\boldsymbol{P}_{X_1}-\boldsymbol{P}_{Z_{2|1}})\boldsymbol{y}}{\hat{\sigma}^2_X} + \log(n)(1+k+p_2)\ .\
\]
Every second-order model compared by mBIC has a fixed $\boldsymbol{X}_1$. Using the identity $\boldsymbol{P}_X=\boldsymbol{P}_{X_1}+\boldsymbol{P}_{X_{2|1}}$ and $\hat{\sigma}^2_X=\boldsymbol{y}^T(\boldsymbol{I}-\boldsymbol{P}_{X})\boldsymbol{y}/g$, it follows
\begin{align*}
    \frac{\boldsymbol{y}^T(\boldsymbol{I}-\boldsymbol{P}_{X_1}-\boldsymbol{P}_{Z_{2|1}})\boldsymbol{y}}{\hat{\sigma}^2_X}&=\frac{\boldsymbol{y}^T(\boldsymbol{I}-\boldsymbol{P}_{X}+\boldsymbol{P}_{X}-\boldsymbol{P}_{X_1}-\boldsymbol{P}_{Z_{2|1}})\boldsymbol{y}}{\hat{\sigma}^2_X}\\
    &=\frac{\boldsymbol{y}^T(\boldsymbol{I}-\boldsymbol{P}_{X})\boldsymbol{y} + \boldsymbol{y}^T(\boldsymbol{P}_{X}-\boldsymbol{P}_{X_1}-\boldsymbol{P}_{Z_{2|1}})\boldsymbol{y}}{\hat{\sigma}^2_X}\\
    &=g+[\text{r}(\boldsymbol{X}_{2|1})-p_2]\frac{\boldsymbol{y}^T(\boldsymbol{P}_{X_{2|1}}-\boldsymbol{P}_{Z_{2|1}})\boldsymbol{y}/[(\text{r}(\boldsymbol{X}_{2|1})-p_2)]}{\hat{\sigma}^2_X}\\
    &=g+[\text{r}[\boldsymbol{X}_{2|1})-p_2]F_{Z_{2|1}}\ .\
\end{align*}    
Therefore we have (after rearranging terms)
\[
    \text{mBIC}=[\text{r}(\boldsymbol{X}_{2|1})-p_2]F_{Z_{2|1}}+ \log(n)p_2 +g+\log(n)(1+k)\ .\
\]
The last two summands are constant across all considered second-order models and so minimization of mBIC is equivalent to minimizing only the first two summands.

\subsection{ECI Construction Algorithm}

We first present the algorithm where $l=0$. In the coordinate exchange algorithm, if a row in $\boldsymbol{D}_u$ that is paired with one or more rows in $\boldsymbol{D}_r$ is manipulated, the corresponding rows in $\boldsymbol{D}_r$ will also be changed.  This retains the pure-error degrees-of-freedom resulting from $\boldsymbol{D}_r$. We do not enforce the same simultaneous exchange on replicated rows that appear in $\boldsymbol{D}_u$.

Consider a coordinate exchange of a row in $\boldsymbol{D}_u$ from initial design $\boldsymbol{D}$. Let $\chi_j$ denote the set of possible values $d_{ij}$ can take. We mainly consider $\chi_j={\pm1}$ or $\{0,\pm1\}$; if $\chi_j=[-1,1]$ one could use a dense grid of points that fill the space. A coordinate exchange of an element $d_{ij}$ from $\boldsymbol{D}_u$ considers the criterion values for candidate designs resulting from changing $d_{ij}$ to a new element $\tilde{d} \in \chi_j$. Note that a coordinate exchange is a special case of a row exchange (i.e., exchanging row $\boldsymbol{d}_i$ with $\tilde{\boldsymbol{d}} \in \bigotimes \chi_j$. We need an efficient calculation for all components of our proposed criterion: $(\boldsymbol{X}_1^T\boldsymbol{X}_1)^{-1}$, $\boldsymbol{A}$, and $c(\alpha,g)$. 


Let $r_i\geq1$ be the total number of rows we are simultaneously exchanging, with one row in $\boldsymbol{D}_u$ and $r_i-1$ rows from $\boldsymbol{D}_r$. Denote the candidate design by $\widetilde{\boldsymbol{D}}$. Let $\boldsymbol{V}=(\boldsymbol{X}_1^T\boldsymbol{X}_1)^{-1}$ of the initial design and denote the full model matrix for $\widetilde{\boldsymbol{D}}$ by $\widetilde{\boldsymbol{X}}=(\widetilde{\boldsymbol{X}}_1 | \widetilde{\boldsymbol{X}}_2)$. We are interested in efficiently calculating $\widetilde{\boldsymbol{V}}=(\widetilde{\boldsymbol{X}}_1^T\widetilde{\boldsymbol{X}}_1)^{-1}$and the updated $v_j$ terms can be obtained from the lower $k \times k$ diagonal elements of $\widetilde{\boldsymbol{V}}$. 

Let $\boldsymbol{x}_i^T=(1,\boldsymbol{d}_i)$ and similarly define $\tilde{\boldsymbol{x}}$ using $\tilde{\boldsymbol{d}}$. Following \cite{meyer1995coordinate}, we have the representation $\widetilde{\boldsymbol{V}}=(\boldsymbol{V}^{-1}+\boldsymbol{F}_1\boldsymbol{F}_2^T)^{-1}$
where $\boldsymbol{F}_1=\sqrt{r_i}(\tilde{\boldsymbol{x}},-\boldsymbol{x}_i)$ and $\boldsymbol{F}_2=\sqrt{r_i}(\widetilde{\boldsymbol{x}} , \boldsymbol{x}_i)$. Then by \cite{sherman1950adjustment}
\[
\widetilde{\boldsymbol{V}}=\boldsymbol{V}-\boldsymbol{V}\boldsymbol{F}_1(\boldsymbol{I}+\boldsymbol{F}_2^T\boldsymbol{V}\boldsymbol{F}_1)^{-1}\boldsymbol{F}_2^T\boldsymbol{V}_1\ .\
\]
This may be calculated efficiently after inverting the $2 \times 2$ matrix $\boldsymbol{I}+\boldsymbol{F}_2^T\boldsymbol{V}\boldsymbol{F}_1$, which has the expression
\begin{align*}
\boldsymbol{I}+\boldsymbol{F}_2^T\boldsymbol{V}\boldsymbol{F}_1 &= \boldsymbol{I}+r_i\begin{pmatrix} \tilde{\boldsymbol{x}}^T \\ \boldsymbol{x}_i^T \end{pmatrix}\boldsymbol{V}\begin{pmatrix} \tilde{\boldsymbol{x}} & -\boldsymbol{x}_i\end{pmatrix}\\
    &= \begin{pmatrix}1+r_i \tilde{\boldsymbol{x}}^T\boldsymbol{V}\tilde{\boldsymbol{x}} & -r_i \tilde{\boldsymbol{x}}^T\boldsymbol{V}\boldsymbol{x}_i\\ r_i\boldsymbol{x}_i^T\boldsymbol{V}\tilde{\boldsymbol{x}} & 1-r_i\boldsymbol{x}_i^T\boldsymbol{V}\boldsymbol{x}_i \end{pmatrix}\ .\
\end{align*}
Letting $v(\boldsymbol{x}_i,\tilde{\boldsymbol{x}})=\boldsymbol{x}_i^T\boldsymbol{V}\tilde{\boldsymbol{x}}$ and $v(\boldsymbol{x}_i)=v(\boldsymbol{x}_i,\boldsymbol{x}_i)$, it follows
\[
(\boldsymbol{I}+\boldsymbol{F}_2^T\boldsymbol{V}\boldsymbol{F}_1)^{-1} = \frac{1}{[1+r_i v(\tilde{\boldsymbol{x}}][1-r_i v(\boldsymbol{x}_i)]+r_i^2 v(\boldsymbol{x}_i,\tilde{\boldsymbol{x}})} \begin{pmatrix} 1-r_i v(\boldsymbol{x}_i) & r_i v(\boldsymbol{x}_i,\tilde{\boldsymbol{x}})\\ -r_i v(\boldsymbol{x}_i,\tilde{\boldsymbol{x}}) & 1+r_i v(\tilde{\boldsymbol{x}})  \end{pmatrix}\ .\
\]
Note a candidate point, $\tilde{\boldsymbol{x}}$, will have a singular $\widetilde{\boldsymbol{V}}$ when
\[
[1+r_i v(\tilde{\boldsymbol{x}}][1-r_i v(\boldsymbol{x}_i)]+r_i^2 v(\boldsymbol{x}_i,\tilde{\boldsymbol{x}}) = 0\ .\
\]


\cite{Jones_2011} provide update formulas for $\boldsymbol{A}$, but we did not find much computational improvement over straightforward calculation of $\widetilde{\boldsymbol{A}}=\widetilde{\boldsymbol{V}}\widetilde{\boldsymbol{X}}_1^T\widetilde{\boldsymbol{X}}_2$. 
It is also difficult to efficiently calculate changes in the $g$ degrees-of-freedom.  A coordinate exchange will cause $\text{r}(\widetilde{\boldsymbol{X}})\tilde{g}=g,g+1,$ or $g-1$ as we are exchanging either a single row or simultaneously exchanging replicated rows. There are certainly special cases in which this updated value can be determined quickly, but we found it easier and faster to determine $\tilde{g}$ by calculating $\text{r}(\widetilde{\boldsymbol{X}})$ via its QR decomposition.






Following the coordinate exchange, we consider exchanging the rows in $\boldsymbol{D}_r$ with any of the unique rows in 
$\boldsymbol{D}_u$, with replacement. We enumerate all possible such exchanges and evaluate the new components efficiently, when possible. In particular, let $\boldsymbol{X}_{1u}=(\boldsymbol{1} \, | \, \boldsymbol{D}_u)$ and $\boldsymbol{V}_u=(\boldsymbol{X}_{1u}^T\boldsymbol{X}_{1u})^{-1}$.  Then $\widetilde{\boldsymbol{V}}=(\boldsymbol{V}_u^{-1} + \widetilde{\boldsymbol{X}}_{1r}^T\widetilde{\boldsymbol{X}}_{1r})^{-1}$, which can be updated efficiently using \cite{sherman1950adjustment}. If the number of possibilities is too large, we recommend either all exchanges without replacement or taking a large random sample of potential exchanges.











Now we consider the case of $\ell > 0$.  To encourage $\ell$ lack-of-fit degrees-of-freedom, we first calculate the number of pure error degrees-of-freedom of the candidate design, $\tilde{p}$, and calculate
$\tilde{\ell}=\tilde{g}-\tilde{p}$. If $\tilde{\ell} \geq \ell$, the design has met the lack-of-fit degrees-of-freedom requirement. Otherwise we need to calculate $$\widetilde{\boldsymbol{C}}_{2|1}=\widetilde{\boldsymbol{X}}_2^T(\boldsymbol{I}-\widetilde{\boldsymbol{X}}_1\widetilde{\boldsymbol{V}}\widetilde{\boldsymbol{X}}_1)\widetilde{\boldsymbol{X}}_2=\widetilde{\boldsymbol{X}}_2^T\widetilde{\boldsymbol{X}}_2 -\widetilde{\boldsymbol{A}}^T\widetilde{\boldsymbol{X}}_1^T\widetilde{\boldsymbol{X}}_1\widetilde{\boldsymbol{A}}\ ,\ $$
which requires no additional matrix inversion. Finally, we calculate and sum the $\ell^*=\ell - \tilde{\ell}$ smallest, positive eigenvalues of $\widetilde{\boldsymbol{C}}_{2|1}$ to update $c(\alpha,g,\tau)$.



\subsection{Simulation study for all-subsets versus guided subsets}
We now showcase our analysis approach with a short simulation study based on that done in \cite{jones2017effective} for ADSDs. We compared the all-subsets mBIC analysis to two versions of the guided subsets analysis, one with a maximal model size of $\lfloor r(\boldsymbol{X}_2)/2 \rfloor$ (as recommended by the \cite{jones2017effective}) and $r(\boldsymbol{X}_2)-1$. We focus on ADSDs with $k=6$ and $k=8$ and $f=2$, which gives two lack-of-fit degrees-of-freedom. All active effects were generated from an exponential distribution with mean $1$ with lower bound set to $2.5$, assigning positive and negative signs at random. The active main effects were assigned at random and all active second-order terms followed the strong heredity principle. We set $\sigma^2=1$ and performed each case $100$ times. The cases are described in Table~\ref{tab:chap2_sect4_cases}.

\begin{table}[ht]
    \centering
    \begin{tabular}{c|c|c|c|c|c}
     Case & $k$ & $n$ & \# Active Main & \# Active 2FIs & \# Active Quad\\ \hline
        1 & 6 & 17 & 4 & 2 & 1\\
        2 & 6 & 17 & 4 & 4 & 3\\ \hline
        3 & 8 & 21 & 6 & 3 & 2\\
        4 & 8 & 21 & 6 & 5 & 3\\ \hline
    \end{tabular}
    \caption{Simulation study cases considered comparing all-subsets analysis to guided subsets.}
    \label{tab:chap2_sect4_cases}
\end{table}

\begin{figure}[ht]
    \centering
    \includegraphics[width=0.75\textwidth]{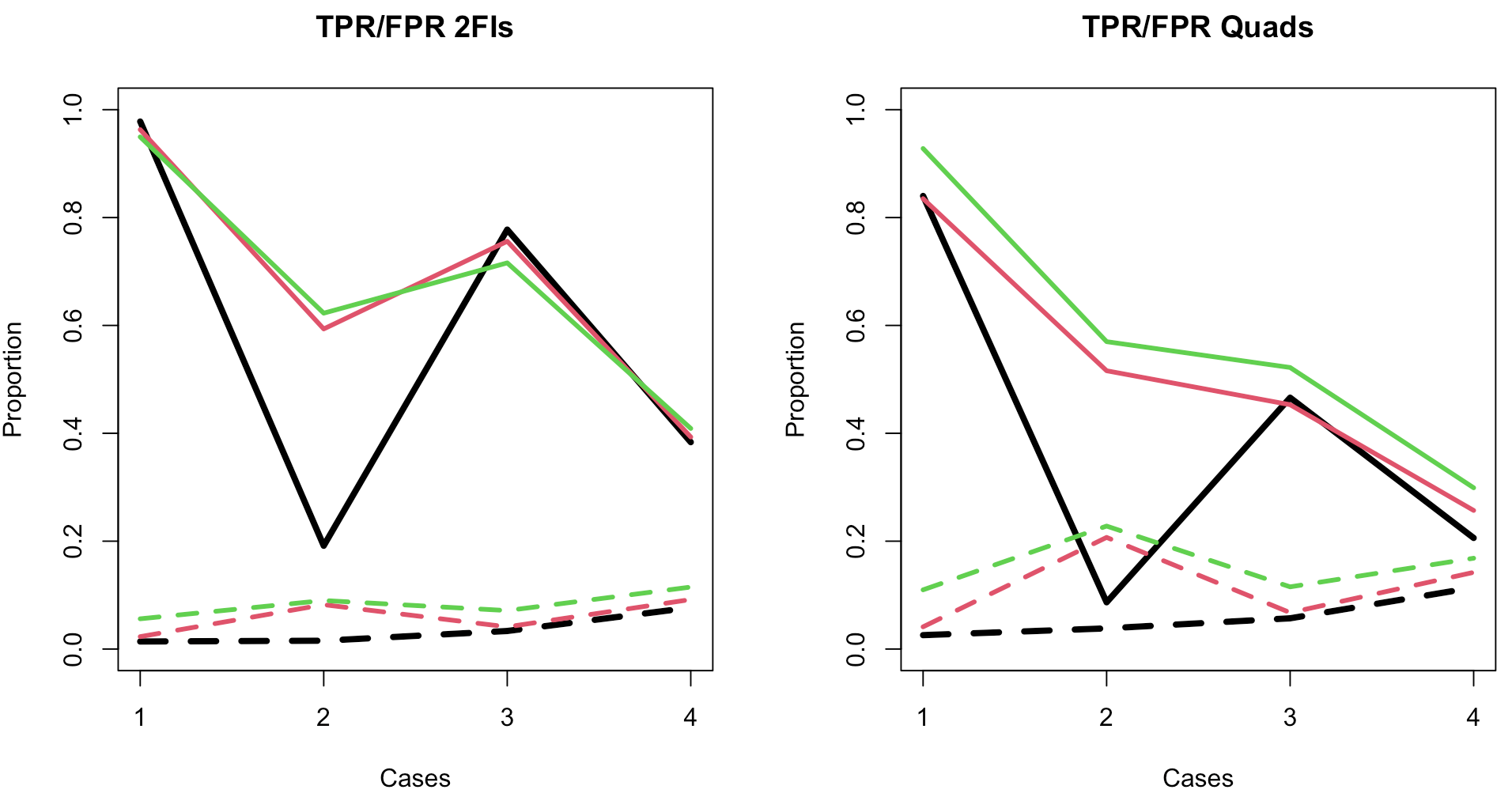}
    \caption{The left and right plots show the $TPR_{2FI}/FPR_{2FI}$ and $TPR_{Q}/FPR_{Q}$, respectively, for the four cases and different analysis methods. The black and red lines correspond to guided subsets with maximal model sizes of $\text{r}(\boldsymbol{X}_2)/2$ and $\text{r}(\boldsymbol{X}_2)-1$, respectively, while the green line is all-subsets. Solid lines give the $TPR$'s and dashed lines are the $FPR$'s. The largest standard error would occur at a proportion of $0.5$, being $\sqrt{0.5^2/1000}=0.0158$.}
    \label{fig:chap2_sect4_results}
\end{figure}

We summarize the simulation study with the following metrics:
\begin{enumerate}
    \item $TPR_{2FI}$ and $FPR_{2FI}$: the proportion of times an active (inactive) two-factor interaction was declared active (inactive);
    \item $TPR_{Q}$ and $FPR_Q$: the proportion of times an active (inactive) quadratic effect was declared active (inactive).
\end{enumerate}
We do not report the true/false positive rate for the main effect analysis because this stage is the same across the three analysis protocols.
The results of the simulation study are shown in Figure~\ref{fig:chap2_sect4_results}. As expected, guided subsets with a maximal model size of $\lfloor r(\boldsymbol{X}_2)/2 \rfloor$ had the lowest false positive rates while all-subsets had the highest false positive rates. For case 1, all three analyses perform comparably for selecting the two-factor interactions, while the all subsets approach increased $TPR_{Q}$ from $0.840$ to $0.928$. All remaining cases involved more second-order terms, leading to lower true positive rates. In case 2, all-subsets had higher $TPR_{2FI}$ and $TPR_{Q}$. For case 3, both guided subsets had slightly higher $TPR_{2FI}$, while the all-subsets approach had the highest $TPR_{Q}$ with a value of $0.522$. Finally, all-subsets had the highest $TPR_{2FI}$ and $TPR_{Q}$ for case 4.

\bibliographystyle{asa}
\bibliography{main}

\begin{thebibliography}{36}
\newcommand{\enquote}[1]{``#1''}
\expandafter\ifx\csname natexlab\endcsname\relax\def\natexlab#1{#1}\fi

\bibitem[{Allen-Moyer and Stallrich(2021)}]{allen2021incorporating}
Allen-Moyer, K. and Stallrich, J. (2021), \enquote{Incorporating {M}inimum
  {V}ariances into {W}eighted {O}ptimality {C}riteria,} \textit{The American
  Statistician}, 76, 1--8.

\bibitem[{Box et~al.(1978)Box, Hunter, and Hunter}]{box1978statistics}
Box, G., Hunter, W., and Hunter, J. (1978), \textit{Statistics for
  {E}xperimenters: {A}n {I}ntroduction to {D}esign, {D}ata {A}nalysis, and
  {M}odel {B}uilding}, Wiley Series in Probability and Statistics - Applied
  Probability and Statistics Section Series, Wiley.

\bibitem[{Box and Meyer(1986)}]{Box1986}
Box, G.~E. and Meyer, R.~D. (1986), \enquote{An {A}nalysis for {U}nreplicated
  {F}ractional {F}actorials,} \textit{Technometrics}, 28, 11--18.

\bibitem[{Cheng and Tang(2005)}]{Min_G_ab}
Cheng, C.~S. and Tang, B. (2005), \enquote{A {G}eneral {T}heory of {M}inimum
  {A}berration and {I}ts {A}pplications,} \textit{The Annals of Statistics},
  33, 944--958.

\bibitem[{Cox and Reid(2000)}]{cox2000theory}
Cox, D.~R. and Reid, N. (2000), \textit{The {T}heory of the {D}esign of
  {E}xperiments}, Chapman and Hall/CRC.

\bibitem[{Deng and Tang(1999)}]{GMA}
Deng, L.~Y. and Tang, B. (1999), \enquote{Generalized resolution and minimum
  aberration criteria for {P}lackett-{B}urman and other nonregular factorial
  designs,} \textit{Statistica Sinica}, 9, 1071--1082.

\bibitem[{Draper and Guttman(1992)}]{draper1992treating}
Draper, N.~R. and Guttman, I. (1992), \enquote{Treating bias as variance for
  experimental design purposes,} \textit{Annals of the Institute of Statistical
  Mathematics}, 44, 659--671.

\bibitem[{DuMouchel and Jones(1994)}]{dumouchel1994simple}
DuMouchel, W. and Jones, B. (1994), \enquote{A {S}imple {B}ayesian
  {M}odification of {D}-{O}ptimal {D}esigns to {R}educe {D}ependence on an
  {A}ssumed {M}odel,} \textit{Technometrics}, 36, 37--47.

\bibitem[{Efroymson(1960)}]{Efroymson}
Efroymson, M. (1960), \enquote{Multiple {R}egression {A}nalysis,}
  \textit{Mathematical Methods for Digital Computers}, 191--203.

\bibitem[{Gilmour and Trinca(2012)}]{gilmour2012optimum}
Gilmour, S.~G. and Trinca, L.~A. (2012), \enquote{Optimum design of experiments
  for statistical inference,} \textit{Journal of the Royal Statistical Society:
  Series C (Applied Statistics)}, 61, 345--401.

\bibitem[{Hastie et~al.(2009)Hastie, Tibshirani, and Friedman}]{ESLtextbook}
Hastie, T., Tibshirani, R., and Friedman, J. (2009), \textit{The Elements of
  Statistical Learning}, New York: Springer, chap. Model Assessment and
  Selection.

\bibitem[{Hofmann et~al.(2021)Hofmann, Gatu, Kontoghiorghes, Colubi, and
  Zeileis}]{hofmann2020lmsubsets}
Hofmann, M., Gatu, C., Kontoghiorghes, E.~J., Colubi, A., and Zeileis, A.
  (2021), \textit{{lmSubsets}: Exact Variable-Subset Selection in Linear
  Regression}, r package version 0.5-2.

\bibitem[{Hurvich and Tsai(1989)}]{AICc}
Hurvich, C.~M. and Tsai, C.~L. (1989), \enquote{Regression and time series
  model selection in small samples,} \textit{Biometrika}, 76, 297--307.

\bibitem[{Jones et~al.(2020)Jones, Hunter, and Montgomery}]{jones2020partial}
Jones, B., Hunter, J.~S., and Montgomery, D.~C. (2020), \enquote{Partial
  {R}eplication of {D}efinitive {S}creening {D}esigns,} \textit{Quality
  Engineering}, 32, 4--9.

\bibitem[{Jones and Nachtsheim(2011{\natexlab{a}})}]{jones2011class}
Jones, B. and Nachtsheim, C.~J. (2011{\natexlab{a}}), \enquote{A {C}lass of
  {T}hree-level {D}esigns for {D}efinitive {S}creening in the {P}resence of
  {S}econd-{O}rder {E}ffects,} \textit{Journal of Quality Technology}, 43,
  1--15.

\bibitem[{Jones and Nachtsheim(2011{\natexlab{b}})}]{Jones_2011}
--- (2011{\natexlab{b}}), \enquote{Efficient {D}esigns with {M}inimal
  {A}liasing,} \textit{Technometrics}, 53, 62--71.

\bibitem[{Jones and Nachtsheim(2013)}]{jones2013definitive}
--- (2013), \enquote{Definitive {S}creening {D}esigns with {A}dded
  {T}wo-{L}evel {C}ategorical {F}actors,} \textit{Journal of Quality
  Technology}, 45, 121--129.

\bibitem[{Jones and Nachtsheim(2017)}]{jones2017effective}
--- (2017), \enquote{Effective {D}esign-{B}ased {M}odel {S}election for
  {D}efinitive {S}creening {D}esigns,} \textit{Technometrics}, 59, 319--329.

\bibitem[{Jones et~al.(2007)Jones, Li, Nachtsheim, and Kenny}]{JLN}
Jones, B.~A., Li, W., Nachtsheim, C.~J., and Kenny, Q.~Y. (2007),
  \enquote{Model {D}iscrimination—{A}nother {P}erspective on {M}odel-{R}obust
  {D}esigns,} \textit{Journal of Statistical Planning and Inference}, 137,
  1576--1583.

\bibitem[{Kiefer(1974)}]{Kiefer}
Kiefer, J. (1974), \enquote{General {E}quivalence {T}heory for {O}ptimum
  {D}esigns ({A}pproximate {T}heory),} \textit{The Annals of Statistics}, 2,
  849--879.

\bibitem[{Leonard and Edwards(2017)}]{leonard2017bayesian}
Leonard, R.~D. and Edwards, D.~J. (2017), \enquote{Bayesian {D}-optimal
  screening experiments with partial replication,} \textit{Computational
  Statistics \& Data Analysis}, 115, 79--90.

\bibitem[{Li and Nachtsheim(2000)}]{MRFD}
Li, W. and Nachtsheim, C.~J. (2000), \enquote{Model-{R}obust {F}actorial
  {D}esigns,} \textit{Technometrics}, 42, 345--352.

\bibitem[{Lu et~al.(2011)Lu, Anderson-Cook, and Robinson}]{Lu2011}
Lu, L., Anderson-Cook, C.~M., and Robinson, T.~J. (2011), \enquote{Optimization
  of {D}esigned {E}xperiments {B}ased on {M}ultiple {C}riteria {U}tilizing a
  {P}areto {F}rontier,} \textit{Technometrics}, 53, 353--365.

\bibitem[{Marley and Woods(2010)}]{marley2010comparison}
Marley, C.~J. and Woods, D.~C. (2010), \enquote{A {C}omparison of {D}esign and
  {M}odel {S}election {M}ethods for {S}upersaturated {E}xperiments,}
  \textit{Computational Statistics \& Data Analysis}, 54, 3158--3167.

\bibitem[{Mead et~al.(2012)Mead, Gilmour, and Mead}]{mead2012statistical}
Mead, R., Gilmour, S.~G., and Mead, A. (2012), \textit{Statistical {P}rinciples
  for the {D}esign of {E}xperiments: {A}pplications to {R}eal {E}xperiments},
  vol.~36, Cambridge University Press.

\bibitem[{Meyer and Nachtsheim(1995)}]{meyer1995coordinate}
Meyer, R.~K. and Nachtsheim, C.~J. (1995), \enquote{The {C}oordinate-{E}xchange
  {A}lgorithm for {C}onstructing {E}xact {O}ptimal {E}xperimental {D}esigns,}
  \textit{Technometrics}, 37, 60--69.

\bibitem[{Montepiedra and Fedorov(1997)}]{MONTEPIEDRA199797}
Montepiedra, G. and Fedorov, V.~V. (1997), \enquote{Minimum bias designs with
  constraints,} \textit{Journal of Statistical Planning and Inference}, 63,
  97--111.

\bibitem[{Pukelsheim(2006)}]{pukelsheim2006optimal}
Pukelsheim, F. (2006), \textit{Optimal {D}esign of {E}xperiments}, SIAM.

\bibitem[{Schwarz(1978)}]{BIC}
Schwarz, G. (1978), \enquote{Estimating the {D}imension of a {M}odel,}
  \textit{The Annals of Statistics}, 6, 461--464.

\bibitem[{Sherman and Morrison(1950)}]{sherman1950adjustment}
Sherman, J. and Morrison, W.~J. (1950), \enquote{Adjustment of an {I}nverse
  {M}atrix {C}orresponding to a {C}hange in {O}ne {E}lement of a {G}iven
  {M}atrix,} \textit{The Annals of Mathematical Statistics}, 21, 124--127.

\bibitem[{Sun(1993)}]{Sun}
Sun, D. (1993), \enquote{Estimation capacity and related topics in experimental
  design,} Ph.D. thesis, University of Waterloo.

\bibitem[{Tang and Deng(1999)}]{G2}
Tang, B. and Deng, L.-Y. (1999), \enquote{Minimum {G$_2$}-aberration for
  nonregular fractional factorial designs,} \textit{Annals of Statistics}, 27,
  1914--1926.

\bibitem[{Tsai et~al.(2007)Tsai, Gilmour, and Mead}]{qb}
Tsai, P.-W., Gilmour, S.~G., and Mead, R. (2007), \enquote{Three-{L}evel
  {M}ain-{E}ffects {D}esigns {E}xploiting {P}rior {I}nformation {A}bout {M}odel
  {U}ncertainty,} \textit{Journal of Statistical Planning and Inference}, 137,
  619--627.

\bibitem[{Weese et~al.(2020)Weese, Stallrich, Smucker, and
  Edwards}]{weese2020strategies}
Weese, M.~L., Stallrich, J.~W., Smucker, B.~J., and Edwards, D.~J. (2020),
  \enquote{Strategies for {S}upersaturated {S}creening: {G}roup {O}rthogonal
  and {C}onstrained {V}ar (s) {D}esigns,} \textit{Technometrics}, 63, 1--13.

\bibitem[{Weese et~al.(2021)Weese, Stallrich, Smucker, and
  Edwards}]{weese2021strategies}
--- (2021), \enquote{Strategies for {S}upersaturated {S}creening: {G}roup
  {O}rthogonal and {C}onstrained {V}ar (s) {D}esigns,} \textit{Technometrics},
  63, 443--455.

\bibitem[{Xiao et~al.(2012)Xiao, Lin, and Bai}]{xiao2012}
Xiao, L., Lin, D. K.~J., and Bai, F. (2012), \enquote{Constructing {D}efinitive
  {S}creening {D}esigns {U}sing {C}onference {M}atrices,} \textit{Journal of
  Quality Technology}, 44, 2--8.

\end{thebibliography}

\end{document}